\documentclass[12pt]{article}  

\usepackage{lmodern} 
\usepackage{amsmath,amssymb,amsthm}
\usepackage{graphicx}
\graphicspath{{graphics/}}
\usepackage{enumitem}
\usepackage{hyperref} 
\usepackage{booktabs} 
\usepackage{natbib}
\usepackage{comment}
\usepackage{bm}
\usepackage{adjustbox}
\usepackage{fix-cm}
\usepackage{lineno} 
\usepackage[letterpaper, top=1in, bottom=1in, left=1in, right=1in, includefoot]{geometry}
\makeatletter
\providecommand\normalsize{%
   \@setfontsize\normalsize{10pt}{12pt}%
   \abovedisplayskip 11\p@ \@plus2\p@ \@minus5\p@
   \abovedisplayshortskip \z@ \@plus3\p@
   \belowdisplayshortskip 6\p@ \@plus3\p@ \@minus3\p@
   \belowdisplayskip \abovedisplayskip
   \let\@listi\@listI}
\makeatother

\theoremstyle{plain} 

\theoremstyle{definition} 

\newtheorem{remark}{Remark}

\theoremstyle{remark} 



\begin{document}


\title{Empirical Best Prediction of Poverty Indicators via Nested Error Regression with High-Dimensional Parameters}

\author{Yuting Chen$^{1,*}$, Partha Lahiri$^2$, Nicola Salvati$^3$}

\date{
$^1$Department of Mathematics \& Statistics, Eastern Kentucky University, KY, USA\\
$^2$Joint Program in Survey Methodology \& Department of Mathematics, University of Maryland, College Park, MD, USA\\
$^3$Department of Economics and Management, University of Pisa, Pisa, Italy\\[1em]
$^*$Corresponding author: \texttt{yuting.chen@eku.edu}
}

\maketitle

\begin{abstract}
The Nested Error Regression Model with High-Dimensional Parameters (NERHDP) is extended to address challenges in small area poverty estimation. A robust and flexible framework is proposed to derive empirical best predictors (EBPs) of small area poverty indicators while accommodating heterogeneity in regression coefficients and sampling variances across areas. To mitigate the computational limitations of the existing algorithm, an efficient estimation procedure is introduced, substantially reducing computation time and enhancing scalability for large datasets. A novel approach for generating area-specific poverty estimates in out-of-sample areas is also developed, improving the reliability of synthetic estimates. Uncertainty is quantified through a parametric bootstrap method specifically tailored to the extended model. 
Under heterogeneous data-generating scenarios, the proposed method yields lower relative bias and relative root mean squared prediction error than existing approaches.
The methodology is further illustrated using data from the 2002 Albania Living Standards Measurement Survey, combined with auxiliary information from the 2001 census, to estimate poverty indicators for 374 municipalities.
\end{abstract}


\section{Introduction}
The United Nations emphasizes the priority of eradicating all forms of poverty by making it the first of its 17 Sustainable Development Goals. Measuring poverty levels across regions is essential for designing effective poverty reduction strategies and ensuring equitable allocation of resources. Accurate information about people's living conditions at a regional level is a basic instrument for targeting policies and programs to reduce poverty. However, progress toward this goal is often slowed by a consistent lack of data on important economic indicators, especially for small population groups and specific geographic areas in developing countries. One major reason for the shortage of disaggregated poverty data is that large scale sample surveys are typically designed to produce reliable estimates of various characteristics of interest for large geographic areas or large population subgroups, rather than for smaller, more granular domains \citep{chatterjee2008parametric}.

Standard survey estimates for national and major subgroups are referred to as `design-based' or `direct' estimates, as they rely solely on the survey data and the selection probabilities specific to the subgroup of interest. The direct (or design-based) estimates become problematic when the sample sizes in the subgroups of interest are small (or even zero). In this situation, model-based methods are increasingly being used to produce what are termed as `small area' or `indirect' estimates, which `borrow strength' across areas through linking models based on auxiliary information coming from censuses or
administrative registers; for a detailed overview of these techniques, refer to \cite{jiang2007linear} and \cite{rao2015small}.

Many measures of poverty and inequality are nonlinear functions of the small area distribution of the economic variable underpinning poverty, usually income or a surrogate such as expenditure. In this paper, we focus on estimating a class of poverty measures known as the Foster–Greer–Thorbecke (FGT) poverty measures, introduced by \cite{foster1984class}. Consider a finite population of size $N$ partitioned into $m$ small areas of size $N_1, \cdots, N_m$. Let $E_{ij}$ be a suitable quantitative measure of welfare for individual $j$ in small area $i$, such as income or expenditure, and let $z$ be a fixed poverty line. This poverty line $z$ serves as the threshold below which an individual is considered to be living in poverty. The FGT family of poverty measures for each small area $i$ is then defined as the area-specific mean
\begin{equation}
    F_{\alpha i} = \frac{1}{N_i}\sum_{j=1}^{N_i} F_{\alpha ij},\; i = 1, \cdots, m,
    \label{FGT_i}
\end{equation}
of the values $F_{\alpha ij}$ defined as
\begin{eqnarray}
    F_{\alpha ij} = \bigg(\frac{z - E_{ij}}{z}\bigg)^{\alpha}I(E_{ij} < z),\; j = 1, \cdots, N_i;\; \alpha = 0, 1, 2,
    \label{FGT_ij}
\end{eqnarray}
where $I(E_{ij} < z) = 1$ if $E_{ij} < z$, otherwise, it equals to 0. When $\alpha = 0$, the FGT measure represents the proportion of individuals living below the poverty line in small area $i$, commonly referred to as the poverty incidence or headcount ratio. For $\alpha = 1$, the measure, known as the poverty gap, captures the average relative distance of individuals to the poverty line within the area. When $\alpha = 2$, the measure, termed poverty severity, highlights the inequality among the poor, with higher values of $F_{2i}$ indicating areas experiencing more severe levels of poverty.

The small area estimation (SAE) literature for poverty measures was revolutionized by the pioneering work of \cite{ell:2003}, which introduced a unit-level approach (ELL) for poverty mapping. This method has become widely adopted for producing fine-grained estimates of FGT measures. 
The ELL approach assumes a two-level model with a common regression relationship between the response and auxiliary variables while allowing for heteroskedastic unit-level errors. Following the framework proposed by \citet{ell:2003}, the unit-level variance components are estimated through a parametric specification of heteroscedasticity as a function of covariates, which may differ from the covariates usued in the mean regression model. In later literature, this variance modeling approach has been referred to as an ``\textit{alpha} model'' (see, e.g., \citealp{nguyen2017small}, \citealp{das2024small}).
Simulated census data are then generated to estimate poverty measures. It is important to note that the ELL approach is a synthetic method that does not rely on a conditional distribution. For further details, we refer readers to their original paper.

Another popular approach for poverty estimation introduced by \cite{molina2010small} is the empirical best prediction (EBP) method through Monte Carlo simulations. This method assumes that the transformed population welfare variables is characterized by a homoskedastic nested error regression (NER) model, as proposed by \cite{battese1988error}. Specifically, it assumes that the regression coefficients and variance components are the same across all small areas, with normally distributed errors. Model parameters are estimated using the maximum likelihood (ML) method or the restricted maximum likelihood (REML) method, and EBPs of poverty indicators are then approximated through random draws from the conditional distribution of the out-of-sample units, given the sample data. \cite{IsabelHB2014} introduced a Bayesian version of this approach, which is particularly useful for quantifying the uncertainty of the estimates. More recently, \cite{morales2021course} (Chapter 10) showed that the conditional expectation can be calculated explicitly and presented closed-form EBPs of the headcount ratio and poverty gap under the NER model.

The effectiveness of such EBP method tends to decline when the number of small areas to be combined becomes large. This limitation arises because the assumption of identical regression coefficients and/or variance components across all small areas, inherent in these models, may not hold universally, leading to potential issues with model misspecification. To address this, random area-specific regression coefficient models \citep{Pra90,hobza2013,rao2015small} and random area-specific sampling variance models \citep{otto1995sampling,arora1997empirical, Kubokawa2016,sugasawa2017} have been suggested in the literature. Such modeling, though useful in some applications, needs more nontrivial assumptions on the joint distribution for a large number of random effects.
On the other hand, fixed effects assumptions on the area specific regression coefficients and sampling variances generally lead to unstable estimates of these fixed effects due to small area specific sample sizes \citep{Jiang2012}.

\cite{lahiri2023nested} introduced a novel approach in the SAE literature, addressing several challenges. Specifically, this approach assumes fixed effects for both area-specific regression coefficients and sampling variances. But they estimate these coefficients using area-specific (with respect to a tuning parameter) estimating equations applied to data from all areas, and then use appropriately constructed residuals for estimation of variance components. When the tuning parameters are known, they have shown that the model parameters can be consistently estimated. This approach, termed as the Nested Error Regression Model with High-Dimensional Parameters (NERHDP), not only offers a more flexible SAE modeling framework but also provides a reliable solution to the challenges of high-dimensional parameter estimation in small area analysis. However, their study focused solely on EBPs for small area means, which are linear functions of the population vector of outcome variable.

In this paper, we extend the NERHDP model to obtain EBPs of poverty indicators, as particular complex and nonlinear FGT measures. Unlike the approach in \cite{molina2010small}, our proposed EBPs account for heterogeneity across small areas in both regression coefficients and sampling variances. Such heterogeneity is common in real-world applications due to variations in socio-economic conditions, data quality, and sampling designs among areas. By extending NERHDP to accommodate FGT poverty measures, we enable the generation of more precise and reliable estimates that are sensitive to area-specific characteristics. With suitable data, this innovation not only enhances the flexibility and applicability of the modeling but also provides a robust methodological framework for addressing the complexities inherent in small area poverty estimation. 

Additionally, we address the computational challenges associated with the algorithm proposed by \cite{lahiri2023nested}, which involves iterative convergence and can be computationally intensive. We introduce a more efficient algorithm that significantly reduces computation time relative to the approach in \cite{lahiri2023nested}, producing estimates within seconds. This improvement ensures the proposed model is not only methodologically robust but also practical for large-scale applications.

Another key feature of our extension is its ability to handle out-of-sample areas, which is particularly important when analyzing a large number of areas. In \cite{molina2010small}, the EBPs become purely synthetic estimates for non-sampled areas due to the absence of observations. In this study, we introduce a new method that enables the estimation of area-specific model parameters even for non-sampled areas. This approach allows for the estimation of poverty measures that are more tailored to specific areas and so less synthetic. While these estimates remain synthetic, they better capture area-specific heterogeneity compared to those in \cite{molina2010small}. Finally, we propose a parametric bootstrap method for quantifying the uncertainty associated with the estimated poverty measures.

In what follows, we focus on two widely used indicators of poverty: the Headcount Ratio (HCR), defined as the proportion of individuals below the poverty line, and the Poverty Gap (PG), which measures the average shortfall of the poor from the poverty line. We first conduct model-based simulation studies to evaluate the performance of our proposed EBP method under the NERHDP model across different data-generating processes. The simulation results show that given suitable data, our method outperforms existing predictors in terms of relative bias and relative root mean squared prediction error, when there is heterogeneity in regression coefficients and/or sampling variances. When regression coefficients are constant across areas, the performance of our proposed method is comparable to that of the EBP introduced in \cite{molina2010small}.

In addition, we demonstrate the benefits of implementing the proposed approach on data from the 2002 Living Standards Measurement Survey (LSMS) in Albania to estimate HCRs and PGs for the 374 municipalities, using the limited auxiliary information available from the 2001 Census. Of these municipalities, $161$ are out-of-sample, meaning they were not included in the original probability sample used for estimation. Consequently, their characteristics were not directly observed in the survey, and any estimates for these areas rely entirely on model-based predictions.

The structure of the paper is as follows. Section \ref{sec:reviewEBP} provides a brief review of the EBP methodology for FGT measures. Section \ref{sec:data} introduces the 2002 Living Standards Measurement Survey (LSMS) in Albania, describes the estimation problem, and outlines the available auxiliary information. Section \ref{sec:NERHD} introduces the best prediction approach under a nested error regression model with high-dimensional parameters, along with the associated new parameter estimation algorithms. In Section \ref{sec:uncertain}, we describe the parametric bootstrap method used to estimate measures of uncertainty. Section \ref{sec:dbsim} presents the results of model-based simulation studies, comparing the performance of the proposed method with existing approaches in terms of bias and variability. In Section \ref{sec:app}, we apply the new methodology to LSMS data to estimate HCR and PG at the municipality level. Finally, Section \ref{sec:discussion} concludes the paper by summarizing the key findings and suggesting directions for future research.

\section{A review of Empirical Best Prediction for FGT measures}\label{sec:reviewEBP}

Like in \cite{molina2010small}, we assume that a random sample of size $n < N$ is drawn from the finite population using a specified sampling design. Let $\Omega$, $s \subset \Omega$, and $r = \Omega - s$ denote the set of indices for all population units, the set of sampled units, and the set of indices for the unsampled units with size $N - n$, respectively. For area $i$, we have the following associated notations: $\Omega_i$, $s_i$, and $n_i$. Note that if area $i$ is not sampled, then $n_i = 0$.

Following the notations from \cite{molina2010small}, we consider a random vector, $\mathbf{y} = (Y_1, \cdots, Y_N)^\prime$, denoting the values of a random variable associated with the $N$ units of a finite population. Let $\mathbf{y}_s$ be the sub-vector of $\mathbf{y}$ corresponding to sample elements $s$ and $\mathbf{y}_r$ be the sub-vector of out-of sample elements $r$. After reordering the units of the population, we can write $\mathbf{y} = (\mathbf{y}_s^\prime, \mathbf{y}_r^\prime)^\prime$. The goal is to predict the value of a real-valued function $\delta = h(\mathbf{y})$ of the random vector $\mathbf{y}$ using the sample data $\mathbf{y}_s$. The best predictor (BP) of $\delta$, which minimizing the mean squared prediction error (MSPE), is given by the conditional expectation
\begin{eqnarray}
    \hat{\delta}^B = \text{E}_{\mathbf{y}_r}(\delta|\mathbf{y}_s),
    \label{BP}
\end{eqnarray}
where the expectation is taken with respect to the conditional distribution of $\mathbf{y_r}$ given the observed sample data $\mathbf{y}_s$ under an assumed model on $\mathbf{y}$.

Suppose there is a one-to-one transformation $Y_{ij} = T(E_{ij})$ of the welfare variables, $E_{ij}$. The vector $\mathbf{y}$ contains the transformed variables $Y_{ij}$ for all population units, following a normal distribution with mean $\bm{\mu}$ and variance $\mathbf{V}$. Using this transformation, the random variable $F_{\alpha ij}$ defined in \eqref{FGT_ij} can be rewritten as:
\begin{equation*}
    F_{\alpha ij} = \bigg(\frac{z - T^{-1}(Y_{ij})}{z}\bigg)^{\alpha}I(T^{-1}(Y_{ij}) < z) =: h_{\alpha}(Y_{ij}),\; j = 1, \cdots, N_i.
\end{equation*}

From \eqref{BP}, the BP of $\delta = F_{\alpha i}$ is given by:
\begin{equation}
       \hat{F}^B_{\alpha i} 
       = \text{E}_{\mathbf{y}_r}(F_{\alpha i}|\mathbf{y}_s) 
       = \frac{1}{N_i}\left\{\sum_{j\in s_i}F_{\alpha ij} + \sum_{j\in r_i}\hat{F}^B_{\alpha ij}\right\},
\end{equation}
where $s_i$ and $r_i$ denote the sets of sample and out-of-sample units belonging to area $i$, respectively, and $\hat{F}^B_{\alpha ij}$ is the BP of $F_{\alpha ij} = h_{\alpha}(Y_{ij})$. The BP of $F_{\alpha ij}$ for $j\in r_i$ is given by:
\begin{eqnarray}
        \hat{F}^B_{\alpha ij} = \text{E}_{\mathbf{y}_r}[h_{\alpha}(Y_{ij})|\mathbf{y}_s] = \int h_{\alpha}(t)f_{Y_{ij}}(t|\mathbf{y}_s)dt,
    \label{eqn:expec_ij}
\end{eqnarray}
where $f_{Y_{ij}}(t|\mathbf{y}_s)$ denotes the conditional density of $Y_{ij}$ given the observed sample data $\mathbf{y}_s$.

Due to the complexity of $h_{\alpha}(t)$, it may not be always feasible to explicitly calculate the expectation in \eqref{eqn:expec_ij}. Assuming that $\mathbf{y} = (\mathbf{y}_s^\prime, \mathbf{y}_r^\prime)^\prime$ follows a multivariate normal distribution with mean vector $\bm{\mu} = (\bm{\mu}_s^\prime, \bm{\mu}_r^\prime)^\prime$ and a covariance matrix partitioned as:
\begin{equation*}
\mathbf{V} = 
    \begin{pmatrix}
   \mathbf{V}_s & \mathbf{V}_{sr}\\
   \mathbf{V}_{rs} & \mathbf{V}_r\\
\end{pmatrix},
\end{equation*}
where $\mathbf{V}_s = \text{var}(\mathbf{y}_s)$, $\mathbf{V}_r = \text{var}(\mathbf{y}_r)$, and $\mathbf{V}_{sr} = \mathbf{V}_{rs}^\prime = \text{cov}(\mathbf{y}_s, \mathbf{y}_r)$.
In the absence of sample selection bias, the conditional distribution of $\mathbf{y}_r$ given $\mathbf{y}_s$ can be expressed as:
\begin{eqnarray}
    \mathbf{y}_r|\mathbf{y}_s \sim \mathcal{N}(\bm{\mu}_{r|s}, \mathbf{V}_{r|s}),
    \label{cond_dist}
\end{eqnarray}
where the conditional mean $\bm{\mu}_{r|s}$ and conditional variance $\mathbf{V}_{r|s}$ are given by:
\begin{equation}\label{cond_mean_var}
    \bm{\mu}_{r|s} = \bm{\mu}_r + \mathbf{V}_{rs}\mathbf{V}^{-1}_s(\mathbf{y}_s - \bm{\mu}_s) \; \text{and} \; \mathbf{V}_{r|s} = \mathbf{V}_r - \mathbf{V}_{rs}\mathbf{V}^{-1}_s\mathbf{V}_{sr}.
\end{equation}

It is important to note that the normality here is adopted as a working modeling assumption to obtain a tractable form for the conditional expectation that defines the empirical best predictor. In applied settings, this assumption may only hold approximately. However, it is standard in EBP methodology and provides a convenient parametric basis for prediction and uncertainty assessment (see, e.g., \citealp{molina2010small}; \citealp{rao2015small}).

\cite{molina2010small} proposed using an empirical approximation of the conditional expectation in \eqref{eqn:expec_ij} through Monte Carlo simulations.
Specifically, a large number $K$ of vectors $\mathbf{y}_r$ are generated from the conditional distribution in \eqref{cond_dist}. Let $Y_{ij}^{(k)}$
denote the $k$th simulated value of the out-of-sample observation $Y_{ij}$ for $j\in r_{i}$, where $k=1, \cdots, K$. The Monte Carlo approximation to the BP for $Y_{ij}$ for $j\in r_{i}$ is then expressed as:
\begin{eqnarray}\label{unitEBP}
    \hat{F}^B_{\alpha ij} = \text{E}_{\mathbf{y}_r}[h_{\alpha}(Y_{ij})|\mathbf{y}_s] \approx \frac{1}{K}\sum_{k = 1}^Kh_{\alpha}(Y_{ij}^{(k)}), \; j \in r_i.
\end{eqnarray}

Since the mean vector $\bm{\mu}$ and covariance matrix $\mathbf{V}$ typically depend on an unknown parameter vector $\bm{\phi}$, the conditional density $f_{Y_{ij}}(t|\mathbf{y}_s)$ also depends on $\bm{\phi}$. Substituting $\bm{\phi}$ with its estimate $\hat{\bm{\phi}}$, we generate simulated values $Y_{ij}^{(k)}$ from the estimated density $f_{Y_{ij}}(t|\mathbf{y}_s, \hat{\bm{\phi}})$. The resulting empirical best predictor (EBP) of the poverty measure $F_{\alpha i}$ is:
\begin{eqnarray*}
    \hat{F}^{EB}_{\alpha i} = \frac{1}{N_i}\left[\sum_{j\in s_i}F_{\alpha ij} + \sum_{j\in r_i}\hat{F}^{EB}_{\alpha ij}\right].
\end{eqnarray*}

Note that the EBP method from \cite{molina2010small} requires linking survey and census households to identify sample indices, which may not be feasible in real-world applications. Often, the survey sample is not a subset of the census. To address this limitation, we follow the approach proposed by \cite{rodas2021pull} and apply Monte Carlo simulations to all population units rather than restricting it to non-sampled units. Thus, the EBP can be expressed as:
\begin{eqnarray}\label{CEB}
    \hat{F}^{EB}_{\alpha i} = \frac{1}{N_i} \sum_j^{N_i}\hat{F}^{EB}_{\alpha ij}.
\end{eqnarray}
For consistency, and without causing confusion, we use \eqref{CEB} as the EBP for the poverty measures throughout this paper. When the sampling fraction $n_i/N_i$ is negligible, which is the most cases in practice, the EBP \eqref{CEB} is practically equivalent to the original formulation proposed by \cite{molina2010small} and further refined by \cite{rodas2021pull} and \cite{molina2022estimation}.

\section{The LSMS and Census data}\label{sec:data}
 The primary dataset used in this study is the 2002 Albania Living Standards Measurement Survey (LSMS), conducted by INSTAT in the Spring of 2002. This survey provides comprehensive insights into the living conditions of the Albanian population, capturing both income and non-income dimensions of poverty. The LSMS sample consists of $3,591$ households, covering a broad range of socio-economic indicators critical for poverty assessment in Albania.
 
The sample design follows a two-stage cluster sampling strategy, with $450$ primary sampling units (PSUs) selected from the 2001 pre-census enumeration areas (EAs). Within each PSU, $8$ households were randomly chosen. The sampling frame was stratified into four main regions to ensure representativeness across Tirana, other urban areas, rural areas, and the three major agro-ecological/economic zones (Coastal, Central, and Mountain). The final dataset comprises of:
 \begin{itemize}
     \item $1,000$ households from the Coastal regions,
     \item $1,000$ households from the Mountain regions,
     \item $991$ households from the Central region,
     \item $600$ households from Tirana.
      \end{itemize}
In the sample, a total of $1,640$ households are rural, and $1,951$ are urban.

The 2002 LSMS data includes observations from 213 of the 374 municipalities in Albania, while the remaining 161 municipalities are entirely out-of-sample, meaning no households were directly surveyed in these areas. Consequently, poverty estimates for these out-of-sample municipalities rely solely on model-based predictions, introducing additional uncertainty and potential extrapolation bias if the characteristics of the out-of-sample areas differ significantly from those of the sampled ones.

Among the observed municipalities, sample sizes vary significantly, ranging from just 6 households in municipalities such as Koder Thu and Hajmel to 600 households in Tirana. The mean sample size is 16.8, with quartiles at 8 (25th percentile), 8 (50th percentile), and 16 (75th percentile). This indicates that several municipalities have very small sample sizes, which hinders the reliability of direct estimates and the stable estimation of random-effect variance components in model-based approaches.

For sampled municipalities, we compute direct estimates of FGT poverty measures, using the weighted average estimator:
\begin{equation*}
    \hat{F}_{\alpha i}^{w} = \frac{1}{\hat{N}_i} \sum_{j \in s_i} w_{ij} F_{\alpha ij},
\end{equation*}
where $w_{ij}$ represents the survey weight for individual $j$ in municipality $i$, and  $\hat{N}_i = \sum_{j \in s_i} w_{ij}$  is a design-unbiased estimator of the population size  $N_i$. The weighted direct estimates and their variances, $\widehat{\text{Var}}_D(\hat{F}_{\alpha i}^{w})$, are computed using the \textit{direct} function from the \texttt{emdi} R package, which employs a nonparametric bootstrap procedure to estimate the design-based variance of the direct estimator $\hat{F}_{\alpha i}^{w}$. This approach is consistent with the general bootstrap methods for complex surveys discussed in the literature (see, e.g., \cite{rao2015small}, Chapter 5).

\begin{figure}[htpb]
    \centering
    \includegraphics[width=0.8\linewidth]{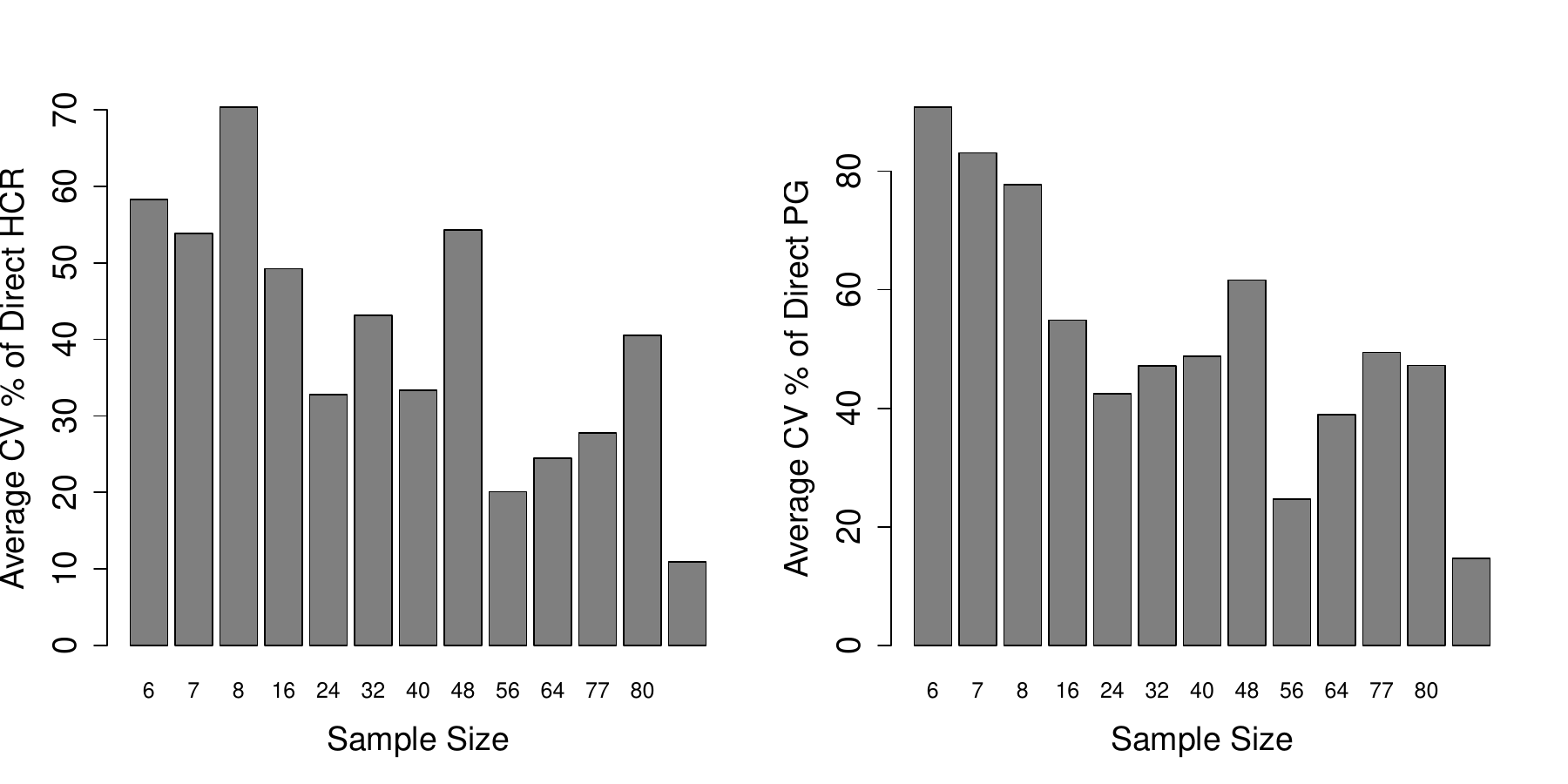}
    \caption{Bar chart of average percent estimated coefficient of variation of direct estimates for HCR and PG versus within-area sample size for all sampled municipalities.}
    \label{fig:cv:size}
\end{figure}

\begin{figure}[htpb]
    \centering
    \includegraphics[width=0.8\linewidth]{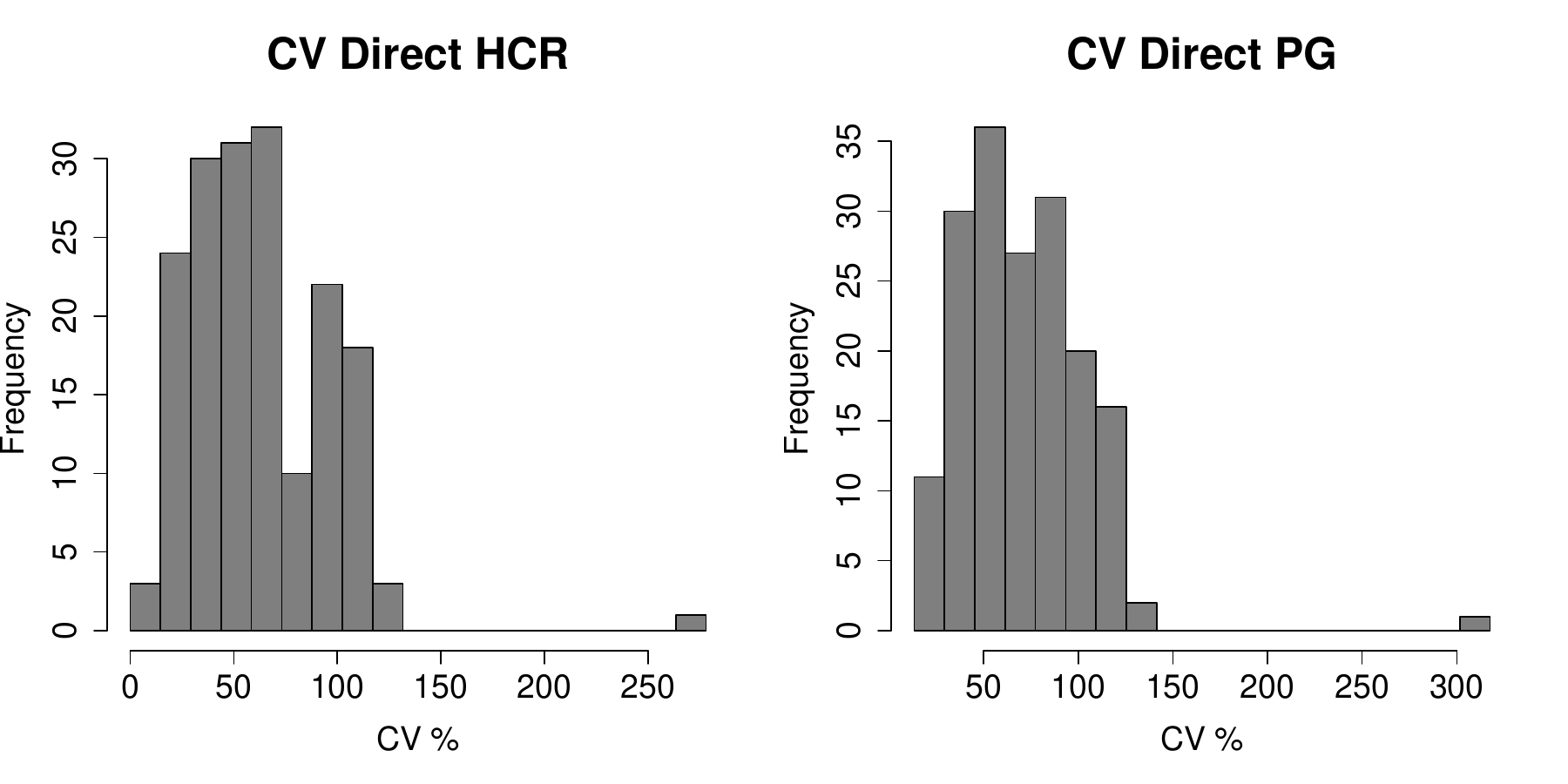}
    \caption{Histograms of percent estimated coefficient of variation of direct estimates for HCR and PG for all sampled municipalities using 2002 LSMS data.}
    \label{fig:cv:direct}
\end{figure}

Figure \ref{fig:cv:size} illustrates the average percent estimated coefficient of variation (CV \%) of direct estimates for HCR and PG, emphasizing the increasing uncertainty in estimates for municipalities with smaller sample sizes. 
Statistical agencies set thresholds for the reliability of estimates based on their CVs. For instance, Statistics Canada considers an estimate to be publishable if its CV is below $16.6\%$, whereas estimates with a CV exceeding $33.3\%$ are deemed unpublishable (\href{https://www150.statcan.gc.ca/n1/pub/89-653-x/2013002/07-eng.htm}{Statistics Canada, 2015}). Similarly, the UK Office for National Statistics (ONS) sets a threshold of 20\% for publication suitability (\href{https://www.ons.gov.uk/employmentandlabourmarket/peopleinwork/employmentandemployeetypes/methodologies/measuringandreportingreliabilityoflabourforcesurveyandannualpopulationsurveyestimates/pdf}{ONS, 2020}). 
If we set the reliability threshold at 20\%, direct survey estimates of the 2002 LSMS dataset would be publishable for only 9 out of the 213 municipalities for HCR and just 2 for PG. Further details are provided in Figure \ref{fig:cv:direct}, which displays the histograms of the estimated percent coefficients of variation for the direct estimates of HCR and PG across all sampled municipalities in the 2002 LSMS data.

This result highlights the limitations of direct estimation methods in small domains, where large sampling variability often renders estimates unreliable. It is important to note that the coefficient of variation is only appropriate for strictly positive outcome variables, such as the poverty indicators considered in this paper (HCR and PG), which are bounded between 0 and 1.  By contrast, variables such as household income or expenditure, which may take non-positive values, require alternative measures of reliability. In such cases, statistical agencies and the SAE literature recommend the use of the standard error (SE) or the relative root mean squared prediction error (RRMSPE) as more suitable indicators of precision (see, e.g., \cite{rao2015small}). Accordingly, in our subsequent simulation studies, we report RRMSPE alongside relative bias to evaluate the performance of predictors.

The application of SAE methods based on the EBP significantly increases the number of municipalities for which reliable estimates can be obtained. This approach requires fitting an appropriate model to the survey data. Estimated model parameters for fixed and random effects are then combined with known population information for each municipality to predict its levels of HCR and PG. The selection of covariates is based on prior poverty assessment studies in Albania, ensuring consistency with previous methodologies. We employ the same household-level covariates as \citet{betti2003poverty} and \citet{tzavidis2008m}, utilizing information available from both the survey and Census. Specifically, the following variables are considered: household size (a strong predictor of poverty), dwelling facilities (e.g., presence of TV, satellite dish, refrigerator, air conditioning, personal computer), ownership variables (e.g., homeownership, land ownership, car ownership). 

 To better understand the distribution of these key covariates, Table \ref{tab:descriptives} presents the proportions for binary variables and the means and medians for continuous variables from both the LSMS and Census data. This provides insights into the variability of the covariates and their potential predictive power for poverty estimation. Examining these statistics helps assess the representativeness of the sample and the suitability of the selected covariates for small area estimation.
\begin{table}[hpbt]
\caption{Descriptive statistics of covariates from LSMS and Census data.}
    \centering
    \begin{tabular}{l|rrr|rrr}
      \hline
      Variable & \multicolumn{3}{c|}{LSMS Data} & \multicolumn{3}{c}{Census Data} \\
               & Median & Mean/Prop  & Std.err. & Median & Mean/Prop & Std.err.\\
      \hline
      Household size & 4 & 4.28 & 1.76 &4 & 4.22 &1.75 \\
      homeownership & & 0.93 &0.26 & &0.93  & 0.25 \\
      land ownership & &0.50 &0.50 & & 0.50 & 0.50 \\
      car ownership & & 0.10 &0.30 & &0.08  & 0.27 \\
      presence of TV & & 0.95 &0.21 & &0.90 & 0.30\\
      satellite dish & &0.21 &0.41 & &0.23 &0.42  \\
      refrigerator & & 0.83 &0.37 & &0.74 &0.44  \\
      air conditioning & &0.02 &0.13 & &0.01 & 0.11 \\
      personal computer & & 0.02 &0.14  & &0.01 & 0.12\\
      \hline
    \end{tabular}
    \label{tab:descriptives}
\end{table}

The descriptive statistics highlight a strong consistency between LSMS and Census data across most covariates, reinforcing the validity of using both sources for small area estimation. Household size exhibits nearly identical distributions, with the same median and very similar mean values, suggesting that the sample adequately represents population characteristics. Similarly, ownership variables such as homeownership and land ownership display minimal differences between the two sources, indicating a high degree of reliability in their measurement. Although minor discrepancies arise in asset ownership, such as car ownership and household appliances, these differences are within an expected range and may be attributed to reporting variations between the survey and the Census. Notably, the proportion of households with a television is slightly higher in LSMS than in the Census ($0.95$ vs. $0.90$), while similar patterns are observed for refrigerator and satellite dish ownership. These variations could reflect differences in reporting behavior or coverage between the two data sources. Overall, the alignment of key covariates between the two datasets provides confidence in their joint use for poverty estimation and underscores the representativeness of the survey data in capturing relevant socioeconomic conditions. It is worth noting that auxiliary variables expressed as proportions that are close to zero or one may have limited predictive value due to their constrained variability. In practice, the contribution of such variables to small area models should be assessed in light of their dispersion across areas.
\section{Nested error regression model with high dimensional parameters}\label{sec:NERHD}
Unlike the traditional NER model used in \cite{molina2010small}, this paper extends the NERHDP model for small area poverty estimation, which accounts for heterogeneity across small areas in both regression coefficients and sampling variances. Specifically,
let $Y_{ij}$ and $\mathbf{x}_{ij}$ denote the values of the study variable and a $p\times 1$ vector of known auxiliary variables for $j$th individual of the $i$th small area. The NERHDP model can be expressed as follows:
\begin{eqnarray}
    Y_{ij} = \beta_{0i} + \mathbf{x}_{ij}^\prime \bm{\beta}_i +\epsilon_{ij},\quad i = 1, \cdots, m; \quad j = 1, \cdots, N_i,
    \label{NERHDP}
\end{eqnarray}
where $\beta_{0i} = \beta_{0} + \gamma_i$. The area effects $\gamma_i \overset{\mathrm{iid}}{\sim} \mathcal{N}(0, \sigma_{\gamma}^2)$ and the errors $\epsilon_{ij} \sim \mathcal{N}(0, \sigma_{\epsilon i}^2)$ are independent. To represent the model at the area level, we define the following vectors and matrices for areas $i = 1, \cdots, m$:
\begin{equation*}
    \mathbf{y}_i = \underset{1\leq j\leq N_i}{\text{col}}(Y_{ij}), \; \bm{\epsilon}_i = \underset{1\leq j\leq N_i}{\text{col}}(\epsilon_{ij}), \; \mathbf{X}_i = \underset{1\leq j\leq N_i}{\text{col}}(\mathbf{x}^\prime_{ij}).
\end{equation*}

Under this model, the response vectors $\mathbf{y}_i$, $i = 1, \cdots, m$, are independent with $\mathbf{y}_i \sim \mathcal{N}(\bm{\mu}_i, \mathbf{V}_i)$, where:
\begin{equation*}
    \bm{\mu}_i = \beta_0\mathbf{1}_{N_i} + \mathbf{X}_i\bm{\beta}_i \;\; \text{and} \;\; \mathbf{V}_i = \sigma_{\gamma}^2\bm{1}_{N_i}\bm{1}_{N_i}^\prime + \sigma_{\epsilon i}^2 \mathbf{I}_{N_i}.
\end{equation*}
Here, $\mathbf{1}_{N_i}$ is an $N_i \times 1$ vector of ones, and $\mathbf{I}_{N_i}$ is the $N_i \times N_i$ identity matrix. Although the area-specific response vectors are conditionally independent under the model, information is shared across areas through the joint estimation of model parameters (see Section~\ref{algoritm_modpara}). In particular, the area-specific regression coefficients are obtained by solving estimating equations that involve data from all sampled areas, and the variance component $\sigma^2_\gamma$ is common across areas. This structure induces borrowing of strength in estimation and prediction, especially for areas with small sample sizes.\par The conditional independence assumption refers to independence given the area-specific random effects and covariates. If additional clustering exists within small areas (e.g., primary sampling units nested within areas) and induces further intra-cluster dependence beyond the area effect, such dependence is not explicitly modeled in the present specification of $\mathbf V_i$. When cluster identifiers are available and such dependence is non-negligible, a multi-level nested-error model incorporating cluster-level random effects may provide a more appropriate specification. 

Following \cite{molina2010small}, for $i = 1, \cdots, m$, we can decompose $\mathbf{y}_i$ into sample and out-of-sample elements $\mathbf{y}_i = (\mathbf{y}^\prime_{is}, \mathbf{y}^\prime_{ir})^\prime$, and the corresponding decomposition on $\mathbf{X}_i$, $\bm{\mu}_i$ and $\mathbf{V}_i$. When the sample size $n_i > 0$, the distribution of $\mathbf{y}_{ir}$ given the sample data is
\begin{eqnarray}
    \mathbf{y}_{ir}|\mathbf{y}_s \sim \mathcal{N}(\bm{\mu}_{ir|s}, \mathbf{V}_{ir|s}),
    \label{cond_dist_i}
\end{eqnarray}
where 
\begin{eqnarray}\label{cond_mean_var_i}
    \bm{\mu}_{ir|s} &=& \beta_0\mathbf{1}_{N_i-n_i} + \mathbf{X}_{ir}\bm{\beta}_i + \sigma_{\gamma}^2\bm{1}_{N_i-n_i}\bm{1}_{n_i}^\prime\mathbf{V}^{-1}_{is}(\mathbf{y}_{is} - \beta_0\mathbf{1}_{n_i} - \mathbf{X}_{is}\bm{\beta}_i) \\
    \mathbf{V}_{ir|s} &=& \sigma_{\gamma}^2(1-B_i)\bm{1}_{N_i-n_i}\bm{1}_{N_i-n_i}^\prime + \sigma_{\epsilon i}^2 \mathbf{I}_{N_i-n_i},
\end{eqnarray}
where $\mathbf{V}_{is} = \sigma_{\gamma}^21_{n_i}1_{n_i}^\prime + \sigma_{\epsilon i}^2 \mathbf{I}_{n_i}$ and $B_i =  \sigma_{\gamma}^2(\sigma_{\gamma}^2 + \sigma_{\epsilon i}^2/n_i)^{-1}$. 

To avoid the computational burden of simulating multivariate normal vectors $\mathbf{y}_{ir}$ of sizes $N_i - n_i,~i = 1, \cdots, m$, the conditional distribution in \eqref{cond_dist_i} can be reformulated using the following model:
\begin{eqnarray}\label{unit_cond_dist}
    \mathbf{y}_{ir} = \bm{\mu}_{ir|s} + u_i\mathbf{1}_{N_i - n_i} + \bm{\epsilon}_{ir},
\end{eqnarray}
where $u_i$ represents new random effects, and $\bm{\epsilon}_{ir}$ denotes a vector of the error terms. These components are independent and satisfy the following distributions:
\begin{eqnarray*}
    u_i \sim \mathcal{N}(0, \sigma_{\gamma}^2(1-B_i)), \quad i = 1, \cdots, m, \quad \text{and} \quad \bm{\epsilon}_{ir} \sim \mathcal{N}(\mathbf{0}_{N_i - n_i}, \sigma_{\epsilon i}^2\mathbf{I}_{N_i - n_i}).
\end{eqnarray*}
This reformulation simplifies the simulation process while maintaining the statistical properties of the original model. As mentioned before, in practice the model parameters $\bm{\phi}_i = (\beta_0, \bm{\beta}^\prime_i, \sigma_{\gamma}^2, \sigma^2_{\epsilon i})$ for $i = 1, \cdots, m$, need to be replaced by suitable estimators $\hat{\bm{\phi}}_i = (\hat{\beta}_0, \hat{\bm{\beta}}^\prime_i, \hat{\sigma}_{\gamma}^2, \hat{\sigma}^2_{\epsilon i})$. 

In line with Chapter 10 of \cite{morales2021course}, when $\alpha = 0$ or $\alpha = 1$ with a shifted logarithmic transformation applied to the welfare variable $E_{ij}$ (i.e., $Y_{ij} = \log(E_{ij}+c)$ for some constant $c$), explicit expressions of the EBPs for the HCR and PG can be obtained. Specifically, the EBP of the HCR for individual $j$ in area $i$ is given by
    \begin{eqnarray}
        \hat{F}^{EB}_{0ij} = \Phi(\hat{\alpha}_{ij}),\quad \hat{\alpha}_{ij} = \hat{v}_{ij|s}^{-1/2}(T(z) - \hat{\mu}_{ij|s}),
    \end{eqnarray}
    where $\Phi(\cdot)$ denotes the cumulative distribution function of the standard normal distribution, $\hat{v}_{ij|s} = \hat{\sigma}_{\gamma}^2(1-\hat{B}_i) + \hat{\sigma}_{\epsilon i}$, and
    \begin{eqnarray*}
        \hat{\mu}_{ij|s} = \hat{\beta}_0 + \mathbf{x}_{ij}^\prime\hat{\bm\beta}_i + \hat{B}_i(\bar{y}_i - \hat{\beta}_0 -\mathbf{\bar x}_{i}^\prime\hat{\bm\beta}_i),
    \end{eqnarray*}
     with $\bar{y}_i = n_i^{-1}\sum_{i=1}^{n_i}y_{ij}$ and $\bar{\mathbf{x}}_i = n_i^{-1}\sum_{i=1}^{n_i}\mathbf{x}_{ij}$. When the shifted logarithmic transformation is applied to $E_{ij}$, the EBP of the PG for individual $j$ in area $i$ becomes
    \begin{eqnarray}
    \hat{F}^{EB}_{1ij} = \frac{z+c}{z}\Phi(\hat{\alpha}_{ij}) - \frac{1}{z}\exp\{\frac{1}{2}\hat{v}_{ij|s} + \hat{\mu}_{ij|s}\}\Phi(\hat{\alpha}_{ij}-\hat{v}_{ij|s}^{1/2}).
    \end{eqnarray}

However, when $\alpha = 2$ or other transformation functions are applied to $E_{ij}$, such as the Box–Cox transformation, explicit expressions for the EBPs of the FGT measures are generally not available. In these cases, Monte Carlo approximations provide a practical solution.

Here, the variables $Y_{ij}$ are generated from the corresponding estimated normal distributions. Given the practical challenges in directly linking the survey sample to the population and under the assumption that the sampling fraction is negligible, for in-sample areas, the elements $\{Y_{ij}, j = 1, \cdots, N_i\}$ are generated from
\[Y_{ij} = \hat{\beta}_0 + \mathbf{x}_{ij}^\prime\hat{\bm{\beta}}_i + \hat{\sigma}_{\gamma}^2\mathbf{1}_{n_i}^\prime\hat{\mathbf{ V}}^{-1}_{is}(\mathbf{y}_{is} - \hat{\beta}_0 \mathbf{1}_{n_i} - \mathbf{X}_{is}\hat{\bm{\beta}}_i) + u_i^* + \epsilon_{ij}^*,\]
where $u_i^* \sim \mathcal{N}(0, \hat{\sigma}_{\gamma}^2(1-\hat{B}_i))$, and $\epsilon_{ij}^* \sim\mathcal{N}(0, \hat{\sigma}^2_{\epsilon i})$.
If an area $i$ is not sampled, then $\{Y_{ij}, j = 1, \cdots, N_i\}$ are generated by bootstrap from 
\[Y_{ij} = \hat{\beta}_0 + \mathbf{x}_{ij}^\prime\hat{\bm{\beta}}_i + \gamma_i^* + \epsilon^*_{ij},\]
where $\gamma_i^* \overset{\mathrm{iid}}{\sim} \mathcal{N}(0, \hat{\sigma}_{\gamma}^2)$ and $\epsilon^*_{ij} \overset{\mathrm{ind}}{\sim} \mathcal{N}(0, \hat{\sigma}^2_{\epsilon i})$. 

Repeat the above generation process $K$ times. Then use the formulas in \eqref{unitEBP} and \eqref{CEB}  obtain the EBP approximation of $F_{\alpha i}$:
\[
\hat{F}^{EB}_{\alpha i} \approx \frac{1}{K}\sum_{k=1}^K\left\{\frac{1}{N_i}\sum_{j}^{N_i}h_{\alpha}(Y_{ij}^{(k)})\right\},
\]
for all areas of interest.

\subsection{Estimation of the high dimensional parameters}\label{algoritm_modpara}
In this study, we introduce a data-driven method for estimating the model parameters $\bm{\phi}_i = (\beta_0, \bm{\beta}^\prime_i, \sigma_{\gamma}^2, \sigma^2_{\epsilon i})$ for $i = 1, \cdots, m$, which effectively leverages data across areas.  
Moreover, our approach offers a significant improvement over \cite{lahiri2023nested} by drastically reducing computational time while maintaining comparable accuracy and performance. These enhancements make our method both more efficient and practical for the analysis of large-scale datasets. 

Throughout the remainder of the paper, we use $\{1, \cdots, m_s\}$ to denote sampled areas and $\{m_s+1, \cdots, m\}$ to index out-of-sample areas, without risk of confusion. The algorithm is detailed in the following steps:
\begin{description}
    \item[Step 1: ] 
    Define $r_{l;i} = q_i^{-1/2}(\mathbf{y}_{ls} - \beta_{0i}\bm{1}_{n_l}-\mathbf{X}_{ls}\bm{\beta}_i), \; l = 1, \cdots, m_s$, where $q_i$ is a scale factor chosen to stabilize the Huber estimating equations and to reduce the influence of outliers (see, e.g. \cite{Bianchi03062015}), ${\mathbf y}_{ls}$ is a $n_l\times 1$ vector of the response variable and $\mathbf{X}_{ls}$ denotes the matrix of dimension $n_l\times p$ individual-level covariates of the sampled units in area $l$. 

Then, obtain the area-specific regression coefficients estimates $\{(\hat{\beta}_{0i}, \hat{\bm{\beta}}_i), i=1, \cdots, m\}$ via solving the following system of equations:
    \begin{equation}
        \sum_{l = 1}^{m_s}\mathbf{X}^\prime_{ls(p+1)}q_i^{1/2}\psi_i(r_{l;i}) = 0,\;i = 1, \cdots, m,
        \label{beta_i}
    \end{equation}
    where $\mathbf{X}_{ls(p+1)}$ is the matrix of dimension $n_l\times(p+1)$ including the intercept term and covariates of the sampled units in area $l$, and $\psi_i(r_{l;i})$ is a $n_l\times 1$ vector obtained from the vector of $r_{l;i}$ with its $j$th element, say $r_{lj;i}$ replaced by $\psi_i(r_{lj;i})$, a chosen known function of $r_{lj;i}$. 
    \begin{remark}
    In this section, we assume that the function $\psi_i$ is fully specified.  For example, consider the form: 
        \begin{equation}
           \psi_i(r) = 2\psi(r)[\tau_iI(r > 0) + (1-\tau_i)I(r\leq 0)], -\infty < r <\infty,  
           \label{psi_i}
        \end{equation}
    where $\psi(r)$ is a known monotone non-decreasing function satisfying $\psi(-\infty) < \psi(0) < \psi(\infty)$, and $\tau_i \in T = (0, 1)$ is a known tuning parameter. Notably, setting $\tau_i = 0.5$ with the identify function for $\psi$ would lead to the standard least square estimator of the regression coefficient vector. 
    
   The tuning parameter $\tau_i$ plays a key role in introducing small area-specific effects into the estimation of both regression coefficients and sampling variances. In contrast to the traditional NER model, which assumes parameter homogeneity across areas, the proposed specification allows $\tau_i$ to vary, thereby inducing heterogeneity in $\bm{\beta}_i$ and error variances $\sigma^2_{\epsilon i}$. This flexibility allows the model to accommodate variations in socio-economic conditions, survey design, and data quality across small areas. Furthermore, when the estimated tuning parameters $\hat{\tau}_i$ are similar across areas, the model naturally simplifies to the conventional NER structure. Consequently, the proposed NERHDP framework attains a desirable balance between flexibility and parsimony, enabling it to adapt to heterogeneous data structures while maintaining stability and strong predictive performance.
    
    Following \cite{chambers2006m}, we assume that $\psi(\cdot)$ is the Huber influence function, with a tuning constant $c = 1.345$, ensuring robust performance in the presence of outliers.
    \end{remark}

    \item[Step 2: ] Define $\tilde{\mathbf{r}}_{l;i} = q_i^{-1/2}(\mathbf{y}_{ls} - \hat{\beta}_{0i}\bm{1}_{n_l}-\mathbf{X}_{ls}\hat{\bm{\beta}}_i)$. We estimate the sampling variance $\{\hat{\sigma}^2_{\epsilon i}, i = 1, \cdots, m\}$ by fitting one-way random effects model to $\{\psi_i(\tilde{\mathbf{r}}_{l;i}), \; l = 1, \cdots , m_s\}$, and obtained the REML estimate of the sampling variance. The model fitting is carried out using the function \texttt{lmer} from the \textsf{R} package \texttt{lme4}.

    \item[Step 3: ] 
    Define $\bar{\mathbf r}^{\star}=\{\bar{r}_i^{\star};\;  i = 1, \cdots, m_s\}^{\prime}$, where $\bar{r}_i^{\star}=n_i^{-1}\sum_{j=1}^{n_i}r_{ij}^{\star}$ and $r_{ij}^{\star}=Y_{ij}-\mathbf{x}_{ij}^{\prime}\hat{\bm{\beta}}_{i}$ with $\hat{\bm{\beta}}_{i}$ from Step 1. We estimate $\beta_0$ by taking the grand residual mean: $\hat{\beta}_0 = 1/n\sum_{i=1}^{m_s}\sum_{j=1}^{n_i}r_{ij}^{\star }$. 
    Define $v_i^{\star} = \sigma^2_{\gamma} + \hat{d}_i$, where $\hat{d}_i = \hat{\sigma}^2_{\epsilon i}/n_i$ for $ i = 1, \cdots, m_s$. Obtain $\hat{\sigma}^2_{\gamma}$  as a solution of the following estimating equation:
    \begin{equation}
    \label{sigmag_i}
        \sum_{i=1}^{m_s} \bigg[\psi^2\bigg((v_i^{\star})^{-1/2} \bar{r}_i^{\star} \bigg)(v_i^{\star})^{-1} -(v_i^{\star})^{-1}w^{\star}\bigg] = 0, 
    \end{equation}
where $w^{\star} = \text{E}[\psi^2(u)]$ with $u$ following a standard normal distribution.
\end{description}

It is worth noting that the proposed procedure shares a common unit-level modeling structure with the approach of \citet{ell:2003}. In both frameworks, poverty indicators are obtained by generating synthetic or predicted unit-level values based on an estimated regression model and associated stochastic components, and then aggregating these values at the small area level.\par However, the conceptual foundations differ. The ELL method constructs synthetic populations using a common regression structure and resampled residual components, without conditioning on observed sample values in the sense of empirical best prediction. By contrast, the NERHDP framework derives empirical best predictors under a nested error regression structure and explicitly exploits the conditional distribution of unobserved units given the observed sample. Furthermore, the proposed model allows regression coefficients and sampling variances to vary across areas through area-specific estimating equations.\par This distinction clarifies that, while both approaches are unit-level and simulation-based, the proposed methodology extends the conditional EBP paradigm to a heterogeneous high-dimensional parameter setting.

\subsection{Definition and estimation of tuning parameter $\tau_i$}\label{subsec:tauX}
Let the subscript $U$ denotes a finite population indicator.
Define the following finite population parameters:
\begin{eqnarray*}
Y_{ij;\tau} &=& \beta_{0\tau;U} + \mathbf{x}_{ij}^\prime\bm{\beta}_{\tau;U};\\
\tau_{ij} &=& \text{argmin}_{\tau\in T}(Y_{ij;\tau} - Y_{ij})^2,
\end{eqnarray*}
for $i = 1, \cdots, m$; $j = 1, \cdots, N_i$.  Note that $Y_{ij;\tau}$ and $\tau_{ij}$ are unknown even for sampled units since $\beta_{0\tau;U}$ and $\bm{\beta}_{\tau;U}$ are unknown finite population parameters.
We are interested in estimating the following finite population parameter: $\tau_i = 1/N_i\sum_{j=1}^{N_i}\tau_{ij}$. The tuning parameter $\tau_i$ plays a pivotal role in the NERHDP framework. It governs the degree of local adaptation of the estimating equations and thus controls the extent to which regression coefficients and sampling variances are allowed to vary across areas. When the $\tau_i$ values are similar across areas, the model approximates the traditional homogeneous NER structure. Conversely, heterogeneous $\tau_i$ values induce area-specific deviations, enabling the model to flexibly accommodate cross-area heterogeneity in regression relationships and dispersion. Therefore, the estimation of $\tau_i$ is not merely a technical component, but a central mechanism through which the proposed method balances flexibility and stability.

For the sampled areas, we use the same data-driven method as \cite{lahiri2023nested} to estimate the tuning parameters for sampled areas, $\tau_i$. Specifically, for a fine grid $\tau\in T = (0, 1)$, we obtain a collection of fitted values for the entire sample:
\begin{eqnarray*}
    \hat{Y}_{ij;\tau} &=& \hat{\beta}_{0\tau} + \mathbf{x}_{ij}^\prime\hat{\bm{\beta}}_{\tau}, \quad i = 1, \cdots, m_s;\quad j=1, \cdots, n_i,
\end{eqnarray*}
using the standard quantile or M-quantile methods, where $\hat{\beta}_{0\tau}$ and $\hat{\bm{\beta}}_{\tau}$ are estimated intercept and regression coefficients, respectively.
For each observation \( Y_{ij} \) in the sample, we identify the fitted line that minimizes the prediction error, defined as the difference between \( Y_{ij} \) and the predicted value from the fitted regression at \( \mathbf{x}_{ij} \). Let \( \hat{\tau}_{ij} \) denote the value of \( \tau \) in the grid corresponding to this best-fitting line. The variability of \( \hat{\tau}_{ij} \) reflects unit-level variability. Provided there are sample observations in area \( i \), and a non-informative sampling method has been used to obtain them, one can consider an estimate of the area-\( i \)-specific tuning parameter is the sample average of \( \hat{\tau}_{ij} \) for that area: $\bar{\hat{\tau}}_i = n_i^{-1}\sum_{j\in s_i}\hat{\tau}_{ij}$. \cite{lahiri2023nested} also introduced an empirical linear best (ELB) predictor for $\tau_i$ denoted as $\hat{\tau}_i^{\text{ELB}}$, which enhances the performance of $\bar{\hat{\tau}}_i$ by leveraging data from all areas. For a detailed explanation, we refer readers to their paper.

For out-of-sample areas with no observations, we propose the following unmatched model to leverage information from observed areas, enabling the estimation of area-specific tuning parameters for completely unobserved regions:
\begin{eqnarray*}
    &&\text{E}(\bar{\hat{\tau}}_i | \tau_i) = \tau_i, \quad \text{Var}(\bar{\hat{\tau}}_i | \tau_i) = \Delta_i; \\
    && \text{E}[\text{logit}(\tau_i)] = \Bar{\mathbf{Z}}_i^\prime\bm{\eta}, \quad i = 1, \ldots, m.
\end{eqnarray*}
We adopt the logit link because $\tau_i \in (0,1)$ by definition and the transformation $\operatorname{logit}(\tau_i)$ maps this constrained parameter to the real line, avoiding boundary issues and stabilizing estimation. Here $\bar{\mathbf Z}_i$ denotes the finite-population (area-level) mean of the auxiliary variables for small area $i$, i.e., the population average of the covariates over area $i$, and $\bm{\eta}$ is the unknown vector of coefficients.

The choice of $\bar{\mathbf{Z}}_i$ is critical in determining the ability of the unmatched model to capture between-area heterogeneity in the tuning parameters $\tau_i$. For out-of-sample areas, where $\tau_i$ cannot be estimated directly from unit-level data, informative auxiliary variables in $\bar{\mathbf{Z}}_i$ that are associated with structural differences in regression coefficients or sampling variances can improve the estimation of $\tau_i$ and enhance predictive accuracy. In contrast, if $\bar{\mathbf{Z}}_i$ has limited explanatory power, the resulting estimates will tend to behave more synthetically.

The variance components $\Delta_i$ are typically assumed to be known; however, in practice, they need to be estimated. In this study, we employ a smoothing technique to estimate $\Delta_i$: 
\[
\hat{\Delta}_i = \frac{1}{n_i} \sum_{i=1}^{m_s} \sum_{j=1}^{n_i} \frac{(\hat{\tau}_{ij} - \bar{\hat{\tau}}_i)^2}{n - m_s}, \quad i = 1, \cdots, m_s.
\]
And we estimate $\bm{\eta}$ by $\hat{\bm\eta}$ that minimizes the following objective function $Q(\bm{\eta})$:
\begin{eqnarray*}
    Q(\bm{\eta}) = \sum_{i= 1}^{m_s} \frac{1}{\hat{\Delta}_i} \left[\bar{\hat{\tau}}_i - \frac{\exp{(\Bar{\mathbf{Z}}_i^\prime\bm{\eta})}}{1+\exp{(\Bar{\mathbf{Z}}_i^\prime\bm{\eta})}}\right]^2.
\end{eqnarray*}
The estimated tuning parameters for the out-of-sample areas are then obtained as follows:
\[
\hat{\tau}_i = \frac{\exp{(\Bar{\mathbf{Z}}_i^\prime\hat{\bm{\eta}})}}{1+\exp{(\Bar{\mathbf{Z}}_i^\prime\hat{\bm{\eta}})}}, \quad i = m_s +1, \cdots, m.
\]

Then, the area-specific model parameters, $\bm{\phi}_i = (\beta_0, \bm{\beta}^\prime_i, \sigma_{\gamma}^2, \sigma^2_{\epsilon i})$ for $i = 1, \cdots, m$, can be estimated using the algorithm outlined in Section \ref{algoritm_modpara}, with the tuning parameters $\tau_i$ replaced by their estimates, $\hat{\tau}_i$, for all small areas.

In the simulation studies presented in Section 6, the adaptive estimation of $\tau_i$ serves as the principal mechanism through which the NERHDP model accommodates heterogeneous data-generating scenarios. In particular, when regression slopes and/or sampling variances vary across areas, the resulting estimates of $\tau_i$ capture this cross-area heterogeneity. This adaptive adjustment is expected to improve predictive performance relative to approaches that impose a common (globally homogeneous) structure across areas.

\section{Uncertainty measures}\label{sec:uncertain}
The assessment of uncertainty of poverty indicators is crucial to analyze the quality of estimates. The MSPE of $\hat{F}^{EB}_{\alpha i}$ is a conventional measure fulfilling this goal, which is given by
\begin{eqnarray}\label{MSE}
\text{MSPE}(\hat{F}^{EB}_{\alpha i}) = E_{\bm{\phi}} (\hat{F}^{EB}_{\alpha i} - F_{\alpha i})^2,
\end{eqnarray}
where $E_{\bm{\phi}}$ denotes expectation with respect to the super-population model \eqref{NERHDP}. However, the analytical approximations to the MSPE are difficult to derive when involving complex parameters such as the FGT poverty measures.

In this paper, we use a parametric bootstrap method that can be applied to produce reasonable estimators of various uncertainty measures, not necessarily MSPE. Our bootstrap scheme builds on the method introduced by \cite{lahiri2023nested}, which is in line with the bootstrap framework introduced by \cite{gonzalez2008bootstrap} and later applied in \cite{molina2010small}, while allowing for a more flexible modeling structure.

The steps of the proposed bootstrap are as follows:
\begin{description}
\item[Step 1: ] Fit model \eqref{NERHDP} to sample data and obtain estimators $\hat{\bm{\phi}}_i = (\hat{\beta}_0, \hat{\bm{\beta}}^\prime_i, \hat{\sigma}_{\gamma}^2, \hat{\sigma}^2_{\epsilon i})$ for $i = 1, \cdots, m$, using the proposed method in section \ref{sec:NERHD};

\item[Step 2: ] Given $\hat{\bm{\phi}}_i$ from Step 1, generate $B$ parametric bootstrap populations using the following model:
\begin{eqnarray}\label{bootsupermodel}
    Y^{*(b)}_{ij} = \hat{\beta}_0 + \mathbf{x}_{ij}^\prime\hat{\bm\beta}_i + \gamma_i^{*(b)} + \epsilon^{*(b)}_{ij},
\end{eqnarray}
where $\gamma_i^{*(b)} \sim \mathcal{N}(0, \hat{\sigma}_{\gamma}^2)$ and $\epsilon^{*(b)}_{ij} \sim \mathcal{N}(0, \hat{\sigma}^2_{\epsilon i})$ are all independently distributed, $i = 1, \cdots, m$, $j = 1, \cdots, N_i$.

\item[Step 3: ] For each population, calculate bootstrap population parameters $F^{*(b)}_{\alpha i} = N_i^{-1}\sum_{j=1}^{N_i}F^{*(b)}_{\alpha ij}$, where $F^{*(b)}_{\alpha ij} = h_{\alpha}(Y^{*(b)}_{ij})$, $b = 1, \cdots, B$.

\item[Step 4: ] Using the same $\hat{\bm{\phi}}_i$ from Step 1, generate bootstrap samples with the same size as the original sample. Specifically, for each bootstrap population $b$, we have the following bootstrap sample:
\begin{eqnarray}\label{bootsupermodel}
    Y^{*(b)}_{ij} = \hat{\beta}_0 + \mathbf{x}_{ij}^\prime\hat{\bm\beta}_i + \gamma_{i}^{*(b)} + \epsilon^{*(b)}_{ij}, i = 1, \cdots, m_s, j = 1, \cdots, n_i,  
\end{eqnarray}
where $\{\gamma_{i}^{*(b)}; i = 1, \cdots, m_s\}$  are selected using the same area indices as in the original sample from $\{\gamma_i^{*(b)}; i = 1, \cdots, m\}$, which are obtained in Step 2. Here $\{\epsilon^{*(b)}_{ij}; i = 1, \cdots, m_s, j = 1, \cdots, n_i\}$ are independently generated from $\mathcal{N}(0, \hat{\sigma}^2_{\epsilon i})$. These errors are not subsets of the population errors generated in Step 2 but are instead independently drawn from the same distributions.

\item[Step 5: ] Fit model \eqref{NERHDP} to bootstrap samples and calculate the bootstrap EBPs, $\hat{F}^{EB*(b)}_{\alpha i}$, $b = 1, \cdots, B$, as described in Sections \ref{sec:reviewEBP} and \ref{sec:NERHD}.

\item[Step 6: ] A Monto Carlo approximation to the theoretical bootstrap estimator $\text{MSPE}_{*}(\hat{F}^{EB*}_{\alpha i}) = E_{\bm{\phi}^*} (\hat{F}^{EB*}_{\alpha i} - F^{*}_{\alpha i})^2$ of $\hat{F}^{EB}_{\alpha i}$ is calculated as
\begin{eqnarray}
    \text{mspe}_{*}(\hat{F}^{EB}_{\alpha i}) = \frac{1}{B}\sum_{b=1}^{B}(\hat{F}^{EB*(b)}_{\alpha i} - F^{*(b)}_{\alpha i})^2.
    \label{bootmse}
\end{eqnarray}
The estimator $\text{mspe}_{*}(\hat{F}^{EB}_{\alpha i})$ is used to estimate $\text{MSPE}(\hat{F}^{EB}_{\alpha i})$ given in \eqref{MSE}.

\end{description}

\section{Model-based Monte Carlo simulations}
\label{sec:dbsim}
We carry out two model-based simulation studies to evaluate the performance of the proposed EBPs of small area HCRs and PGs. In this paper, we followed \citet{Bianchi03062015} in choosing $q_i$. In the model-based simulations, we use the following two auxiliary variables: $x_1$, household size (continuous) and $x_2$, land ownership (binary) from the 2001 Albanian census to repeatedly generate synthetic populations. For the synthetic populations, our small areas are communes and we have $m = 374$ communes with different subpopulation sizes, $N_i$, ranging from 141 to 89764. Then, a sample is drawn from each generated population using a probability sampling design. Additionally, we use different models to generate such synthetic populations for evaluating different predictors under different simulation conditions.

In both simulation studies, we use a simple random sampling from each small area population. Since we focus on predicting small area HCR and PG, the explicit-form of EBP are available. Therefore, we include both the explicit form and Monte Carlo approximation of the EBP under our proposed NERHDP model, which enables us to check the efficiency and variability of Monte Carlo approximation. In addition, we compare our proposed performance to other existing predictors. Specifically, the following predictors of HCR and PG are considered in our model-based simulations:
\begin{description}
    \item[(a) ] Direct estimator (DIRECT),
    \item[(b) ] The Monte Carlo approximation of the EBP based on the traditional NER model from \cite{molina2010small} (MR),
    \item[(c) ] The explicit-form EBP based on the traditional NER model from \cite{morales2021course} (MRE),
    \item[(d) ] The Monte Carlo approximation of EBP based on the proposed NERHDP model (CLS),
    \item[(e) ] The explicit-form EBP based on the proposed NERHDP model (CLSE),
    \item[(f) ] The simplified ELL-type estimator based on the algorithmic structure of \citet{ell:2003}, implemented under the comparative simulation framework adopted in \citet{molina2010small} (hereafter, SELL). 
\end{description}

Note that both MR and MRE are based on the following NER model:
\begin{eqnarray*}
    y_{ij} = \beta_0 + \mathbf{x}_{ij}^\prime \bm{\beta} + \gamma_i + \epsilon_{ij}, \quad i = 1,\dots,m;\; j = 1,\dots,N_i,
    \label{NERMR}
\end{eqnarray*}
where $\bm\beta$ is the vector of constant regression coefficients, the area effects $\gamma_i \sim \mathcal{N}(0, \sigma_{\gamma}^2)$ defined at the domain level, and the errors $\epsilon_{ij} \sim \mathcal{N}(0, \sigma_{\epsilon}^2)$ are independent.

The MR predictor for a poverty indicator $F_{\alpha i}$ is approximated using Monte Carlo simulation:
\begin{equation}
\hat{F}^{EB}_{\alpha i} = \frac{1}{N_i} \sum_{j=1}^{N_i} \frac{1}{K} \sum_{k=1}^K h_\alpha\left(y_{ij}^{*(k)}\right),
\end{equation}
where $y_{ij}^{*(k)}$ is the $k$th simulated value of $y_{ij}$ drawn from its conditional distribution given the sample. In our simulations, we set $K = 100$. The MRE approach, using the explicit analytical expressions for the EBP derived under the NER model \eqref{NERMR}, is described in Chapter 10 of \cite{morales2021course}.

While our simulation setup is similar to \cite{molina2010small} in that clusters are treated as small areas (domains), the simplified ELL-type predictor (SELL) used in our simulation studies adopts the basic algorithmic structure of \citet{ell:2003}, but is implemented under the controlled comparative framework of \citet{molina2010small}. It differs from the usual ELL setup, where clusters are nested within small areas.

Specifically, the underlying model is
\begin{eqnarray}
    Y_{ij} = \beta_0 + \mathbf{x}_{ij}^\prime \bm{\beta} + \gamma_i + \epsilon_{ij}, \quad i = 1,\dots,m;\; j = 1,\dots,N_i,
    \label{NERmod}
\end{eqnarray}
where the unit-level error term $\epsilon_{ij}$ has a flexible variance structure $\sigma^2_{\epsilon ij}$ to accommodate heteroskedasticity.  Initial estimates of the regression coefficients in \eqref{NERmod} are obtained via OLS estimation. Denote the residuals from this regression by $\hat{r}_{ij}$, which are decomposed as
\begin{eqnarray*}
\hat{r}_{ij} = \hat{r}_{i.} + (\hat{r}_{ij} - \hat{r}_{i.}) = \hat{\gamma}_i + \hat{e}_{ij},
\end{eqnarray*}
where the subscript `.’ denotes an average over that index. The terms $\hat{e}_{ij}$ can be used to estimate the variance $\sigma^2_{\epsilon_ {ij}}$ through the \textit{alpha} model, which uses a logit transformation and may involve a rich set of covariates different from $\mathbf{X}$ in the mean regression model.

The $k$th set of synthetic values is then generated for all population units as
\begin{eqnarray*}
y^{syn(k)}_{ij} =\hat{\beta}_0 + \mathbf{x}_{ij}^\prime\hat{\bm\beta} + \hat{\gamma}_i^{(k)} + \hat{e}_{ij}^{*(k)},
\end{eqnarray*}
where $\hat{\gamma}_i^{(k)}$ is sampled with replacement from the set of $\hat{\gamma}_i$, and $\hat{e}_{ij}^{*(k)}$ is sampled with replacement from the set of standardized residuals and then scaled by $\hat{\sigma}_{\epsilon ij}$. The ELL estimator of the poverty indicator $F_{\alpha i}$ is finally computed as
\begin{equation*}
\hat{F}^{ELL}_{\alpha i} = \frac{1}{N_i} \sum_{j=1}^{N_i} \frac{1}{K}\sum_{k=1}^{K}h_\alpha(y^{syn(k)}_{ij}).
\end{equation*}

It is important to clarify that this implementation does not aim to replicate the full applied modeling strategy described in \citet{ell:2003}, which typically involves extensive variable selection, higher-order terms, interaction effects, contextual variables, and possibly cluster-level effects. Instead, consistent with \citet{molina2010small}, we adopt a simplified structure in order to compare prediction mechanisms under a common and controlled data-generating process. This allows us to isolate differences attributable to the prediction methodology rather than to model selection or specification choices. For additional details, we refer readers to \cite{ell:2003}.

\subsection{Model-based simulation I}
In the first simulation study, we assume all 374 communes are selected. Since all areas are sampled in this setting (i.e., $n_i > 0$ for all $i$), the area-specific tuning parameters $\tau_i$ can be estimated directly using within-area sample information, as described in Section \ref{subsec:tauX}.

Within each commune, 3\% of units are selected by simple random sampling without replacement. Consequently, the within-area sample sizes, $n_i$, vary across communes from 4 to 2693. The welfare variable $E_{ij}$ is obtained as an exponential function of $Y_{ij}$. That is, the transformation $T(\cdot)$ is defined as $Y_{ij} = \log(E_{ij})$. The values of $Y_{ij}$ are generated according to the following model:
\begin{eqnarray*}
    Y_{ij} = 9.5 + \beta_{1i}x_{1ij} + \beta_{2i}x_{2ij} + \gamma_i + \epsilon_{ij}, ~i = 1, \cdots, m; ~ j = 1, \cdots, N_i.
\end{eqnarray*}
The slopes $(\beta_{1i}, \beta_{2i})$, the random area effects $\{\gamma_i\}$ and sampling errors $\{\epsilon_{ij}\}$ are independently generated under the following three simulation conditions:
\begin{description}
    \item[(a) Homogeneous model $(0,0):$ ] All areas share the same slopes $(\beta_{1i}, \beta_{2i}) = (-0.3, 0.5)$, $\gamma_i \sim \mathcal{N}(0, 0.1)$, and $\epsilon_{ij} \sim \mathcal{N}(0, 0.5)$. This setup corresponds to the NER model considered in \cite{molina2010small}.
    \item[(b) Varying slopes $(\beta,0):$ ] Areas $i=1,\dots,184$ have $(\beta_{1i}, \beta_{2i}) = (-0.3, 0.5)$, while areas $i=185,\dots,374$ have $(\beta_{1i}, \beta_{2i}) = (-0.5, -0.3)$. Random effects and error terms are generated as in (a).
    \item[(c) Varying slopes and heteroskedasticity $(\beta,\sigma^2_{\epsilon}):$ ] Same slope variation and random effects as in (b), but with $\epsilon_{ij} \sim \mathcal{N}(0, \sigma^2_{\epsilon i})$ where $\sigma^2_{\epsilon i} \sim \text{Gamma}(1, 2)$ independently across areas.
\end{description}

Each scenario is independently simulated 1000 times. The performance of the above predictors is assessed using the following three criteria: 
\begin{description}
    \item[(a) Relative Bias (RB): ]  The average RB across all small areas. For a given area, RB is defined as the ratio of the average difference between the estimated and true small-area FGT measures to the true small-area FGT measure, averaged over simulations.
    \item[(b) Relative Root Mean Squared Prediction Error (RRMSPE): ]  The average RRMSPE across all small areas. For a given area, RRMSPE is the ratio of the square root of the average squared difference between the estimated and true small-area FGT measures to the true measure, averaged over simulations.
    \item[(c) Efficiency (EFF): ]   Defined as the ratio of the RRMSPE of each estimator or predictor to that of the baseline estimator, with MRE serving as the baseline.
\end{description}

\begin{figure}[htbp]
        \centering
        \includegraphics[width=0.8\linewidth]{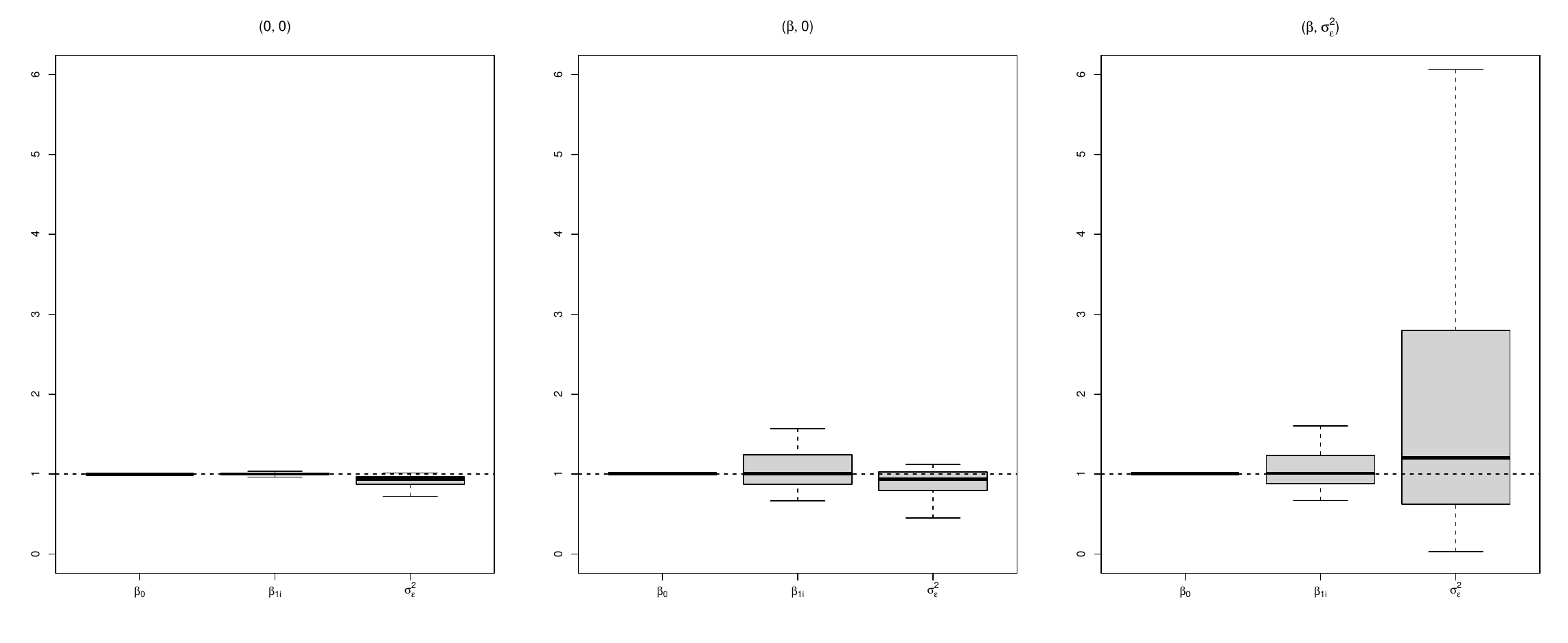}
        \caption{Box-plots displaying ratios of estimates of regression coefficients and area specific sampling variances to their corresponding true values under repeated sampling in our model-based simulation experiment.}
        \label{fig:model para}
    \end{figure}

Figure \ref{fig:model para} presents the performance of our proposed algorithm in estimating the regression coefficients and variance components under the NERHDP model. Across all three simulation scenarios, no evident bias is observed in the estimation of regression coefficients. In scenarios $(0,0)$ and $(\beta,0)$ where the sampling variance is constant across areas, the proposed algorithm exhibits slight underestimation of the area-specific sampling variance.

Table \ref{tab1:all} presents the average values of RB, RRMSPE, and EFF for different simulation scenarios and predictors. The results show that the Monte Carlo approximations MR and CLS generally have slightly larger RB and RRMSPE compared to MRE and CLSE, respectively, but their overall performance is similar. Therefore, in the remainder of this section, we focus on evaluating the Monte Carlo approximations, as they are applicable to more general poverty indicators. According to Table \ref{tab1:all}, our proposed method, CLS, achieves smaller RB than both MR and SELL across all three simulation scenarios. As expected, in scenario $(0,0)$, i.e., the traditional nested error regression model, MR performs best in terms of RRMSPE. In contrast, for the other two scenarios, $(\beta,0)$ and $(\beta,\sigma^2_\epsilon)$ where either the slopes or both the slopes and sampling variances vary across areas, CLS outperforms the other predictors in terms of RRMSPE. 

    \begin{table}[htbp]
    \centering
    \caption{Model-based simulation I results: performance of predictors of small area FGT measures for all sampled areas}
    \begin{tabular}{cccccccccccc}
      \toprule
     Predictor    &\multicolumn{6}{c}{Results (\%) for the following scenarios}\\
     & \multicolumn{3}{c}{Headcount Ratio} & \multicolumn{3}{c}{Poverty Gap}\\
     & (0, 0) &($\beta$, 0) &($\beta$, $\sigma_{\epsilon}^2$) & (0, 0) &($\beta$, 0) &($\beta$, $\sigma_{\epsilon}^2$)\\
     \midrule
     &\multicolumn{6}{c}{Average Relative Bias}\\
    DIRECT & -0.023 & -0.018 &0.018 & -0.025 &-0.008 &-0.007\\
     MR&  2.154 &  14.085 & 18.471 &  3.445 &  22.785 & 23.414  \\
    MRE & 2.032 & 13.929 &18.321  &  3.299 & 22.566 &23.221 \\
     CLS &  0.067 &  3.611 & 7.530 &   -0.648 &  5.939  & 7.297  \\
    CLSE & -0.040 & 3.483 &7.412   & -0.776 & 5.768 & 7.154\\
     SELL &   15.428 &  94.413  & 93.786   &  25.727 &  179.278 &  148.935  \\
         &\multicolumn{6}{c}{Average RRMSPE}\\
    DIRECT &  25.724 &32.028 &32.565 &31.901 &39.374 &40.162 \\
     MR  &  14.543 & 26.096 & 29.419 &  18.875 & 41.540 & 38.479  \\
    MRE & 14.453 &25.955 &29.287  & 19.069 &41.226 &38.197 \\
     CLS &  14.967 & 19.908 & 22.095 &  19.069&  27.943 & 28.318  \\
    CLSE & 14.897  & 19.825 &22.015  & 18.986 &27.813 &28.176 \\
     SELL & 54.253 &  179.599  & 163.272  &  77.682 & 334.468& 242.062  \\
       \hline
    \end{tabular}
    \label{tab1:all}
    \end{table}

\subsection{Model-based simulation II}
In this simulation study, we intentionally exclude 161 areas from the sample, following the sampling pattern of the 2002 Albanian LSMS data, to assess the performance of our proposed EBP methods for out-of-sample areas. Specifically, 213 communes are selected, and this selection is fixed while simple random samples (without replacement) are repeatedly drawn from these areas. Consequently, 161 areas are excluded from the sampling process.

As described in subsection \ref{subsec:tauX}, we propose a model that relates
$\tau_i$ to the auxiliary information $\bar{\bm Z}_i$. However, since $\tau_i$ serves as a tuning parameter, it is difficult to incorporate it directly into the population generation model. Therefore, we assume that $(\beta_{1i}, \beta_{2i})$ are functions of $\bar{\bm Z}_i = (\bar{X}_{1i}, \bar{X}_{2i})$, which may indirectly capture the relationship between $\tau_i$ and $\bar{\bm Z}_i$. In this experiment, we consider the following three data-generation scenarios:
\begin{description}
    \item[(a) Homogeneous model $(0,0):$ ] All areas share the same slopes $(\beta_{1i}, \beta_{2i}) = (-0.3, 0.5)$, $\gamma_i \sim \mathcal{N}(0, 0.1)$, and $\epsilon_{ij} \sim \mathcal{N}(0, 0.5)$.
    \item[(b) Varying slopes $(\beta,0):$ ] $(\beta_{1i}, \beta_{2i})$ are generated through functions of $(\bar{X}_{1i}, \bar{X}_{2i})$:
        $\beta_{1i} = 0.2 \times \bar{X}_{1i}$,
       $ \beta_{2i} = 5 \times \bar{X}_{2i}$,
    where $i = 1, \cdots, 374$, with $\bar{X}_{1i}$ having a mean of 4.45 and standard deviation of 0.526, and $\bar{X}_{2i}$ having a mean of 0.22 and standard deviation of 0.277. The $\gamma_i$ and $\epsilon_{ij}$ are generated as in (a).
    \item[(c) Varying slopes and heteroskedasticity $(\beta,\sigma^2_{\epsilon}):$ ] Same slope variation and random effects as in (b), but with $\epsilon_{ij} \sim \mathcal{N}(0, \sigma^2_{\epsilon i})$ where $\sigma^2_{\epsilon i} \sim \text{Gamma}(1, 2)$ independently across areas.
\end{description}

Tables \ref{tab3:outall} and \ref{tab4:out} report the average values of RB, RRMSE, and EFF for various predictors when out-of-sample areas are involved. While both MR and CLS exhibit better overall performance than SELL, all three predictors show similar performance for out-of-sample areas in the scenario $(0, 0)$, as expected, since both MR and CLS reduce to synthetic predictions in this case. For the scenarios $(\beta, 0)$ and $(\beta, \sigma^2_{\epsilon})$, our proposed CLS method demonstrates the best performance in terms of RB and RRMSE, both overall and for out-of-sample areas. 
    \begin{table}[htbp]
    \centering
    \caption{Performance of predictors of small area FGT measures for all areas when out-of-sample areas are involved.}
    \begin{tabular}{cccccccccccc}
      \toprule
     Predictor    &\multicolumn{6}{c}{Results (\%) for the following scenarios}\\
     & \multicolumn{3}{c}{Headcount Ratio} & \multicolumn{3}{c}{Poverty Gap}\\
     & (0, 0) &($\beta$, 0) &($\beta$, $\sigma_{\epsilon}^2$) & (0, 0) &($\beta$, 0) &($\beta$, $\sigma_{\epsilon}^2$)\\
     \midrule
     &\multicolumn{6}{c}{Average Relative Bias}\\
      {MR}&   8.300 &   26.301 &   26.304&   13.398 &  25.533 &    26.092 \\
    MRE &  8.036 & 25.838 & 25.835 &  13.040 &   24.994 &  25.531\\
      CLS &   7.143 &     6.459  &    6.315 &    10.944  &    8.732   &     8.623  \\
    CLSE & 6.874 &   6.303  & 6.158 &  10.530  & 8.621 & 8.516\\
      SELL &    15.314 &   84.199  &  84.272  &   25.550 &   91.191 &   91.973  \\
         &\multicolumn{6}{c}{Average RRMSPE}\\
      MR  &   32.424 &  36.610 &   37.311 &   45.137 &  40.147 &   42.183 \\
    MRE & 32.176 &36.161 & 36.845 &  44.749 & 39.605 & 41.592\\
      CLS &   32.542 &  14.040 &   13.972&    44.976&   18.037 &   18.176 \\
    CLSE &  32.279 & 13.883  &13.815 &  44.529 &17.919 & 18.026\\
      SELL &  54.209 &   91.022  &  91.338  &    77.59 &   102.120 &  103.529  \\
       \hline
    \end{tabular}
    \label{tab3:outall}
    \end{table}
    
    \begin{table}[htbp]
    \centering
    \caption{Performance of predictors of small area FGT measures for out-of-sample areas.}
    \begin{tabular}{cccccccccccc}
      \toprule
     Predictor    &\multicolumn{6}{c}{Results (\%) for the following scenarios}\\
     & \multicolumn{3}{c}{Headcount Ratio} & \multicolumn{3}{c}{Poverty Gap}\\
     & (0, 0) &($\beta$, 0) &($\beta$, $\sigma_{\epsilon}^2$) & (0, 0) &($\beta$, 0) &($\beta$, $\sigma_{\epsilon}^2$)\\
     \midrule
     &\multicolumn{6}{c}{Average Relative Bias}\\
     {MR}&  17.198 &  80.313 &  84.274&  27.654 & 83.879 &   85.204 \\
    MRE & 16.623 & 78.924 & 82.921 &  26.856 &  82.112  & 83.387\\
     CLS &  16.985 &   21.112  &  20.708 &   27.042 &  25.127  &   25.998  \\
    CLSE & 16.332 &  20.730  & 20.326 &  26.090  &24.699  & 25.568\\
     SELL &   18.496 &  215.009  & 218.421   &  29.732 &  237.951 &  240.449  \\
         &\multicolumn{6}{c}{Average RRMSPE}\\
     MR  &  57.993 & 97.731 &  101.802 &  82.577 &  104.957 &  109.189 \\
    MRE & 57.396 &96.426 & 100.200 & 81.673 &103.165 & 107.287\\
     CLS &   57.842 & 31.468 &  31.643&  82.088&  40.510 &  43.643 \\
    CLSE & 57.192 & 31.083  &31.255 & 81.029 &40.055  & 43.168\\
     SELL & 61.842 &  226.861  & 232.650  &   87.297 &  240.298 & 240.377  \\
       \hline
    \end{tabular}
    \label{tab4:out}
    \end{table}

Overall, our findings should be interpreted within the scope of the data-generating mechanisms examined in the simulation study. Under heterogeneous data-generating scenarios, the results indicate that the proposed method outperforms the existing approaches. In contrast, when the traditional homogeneous NER assumptions are satisfied, the MR/MRE predictors remain competitive and may be preferable due to their more parsimonious structure.

\section{Application: poverty mapping for Albania}\label{sec:app}
Poverty mapping is a crucial tool for understanding the spatial distribution of poverty and guiding the implementation of poverty alleviation programs. In Albania, where the economy is predominantly rural and income is difficult to measure accurately, poverty is assessed using a consumption-based measure \citep{betti2003poverty}. In this section, we use 2002 Living Standards Measurement Survey (LSMS) data to estimate poverty indicators for 374 municipalities in Albania. To supplement this information, we incorporate auxiliary data from the 2001 Census, which covers 726,895 households across Albania. The descriptive statistics of these datasets have been presented in Section \ref{sec:data}, providing an overview of their key characteristics and ensuring their suitability for small area estimation. 

Household consumption expenditure is used as the primary welfare variable, denoted by $E_{ij}$. We apply a logarithmic transformation to obtain $Y_{ij}$ as the response variable. The covariates include household size, dwelling facilities, and ownership variables, based on prior poverty assessment studies in Albania \citep{betti2003poverty,tzavidis2008m}. These covariates are described in detail in Section \ref{sec:data}.

While these covariates were selected to ensure comparability with earlier studies, they are context-specific and may not be directly generalizable to other regions or applications. In general small area estimation (SAE) settings, model specification and covariate selection are critical components that can substantially influence predictive performance and statistical inference. Accordingly, conclusions regarding the relative performance of alternative predictors should be interpreted conditional on the adopted model specification. The poverty line is set at 4,891 Leks per month for all districts, following \cite{betti2003poverty}. 

In our implementation, the area-level covariates $\bar{\mathbf Z}_i$ entering the logit model for $\tau_i$ are computed from the 2001 Census, that is, they represent finite-population means for each municipality. When population-level aggregates are unavailable in other applications, $\bar{\mathbf Z}_i$ may be replaced by design-unbiased estimators derived from the survey data, with appropriate adjustments to account for the additional estimation variability. Alternatively, a two-stage modeling strategy or calibration to external aggregate sources may be employed. We discuss these practical alternatives and their implications for uncertainty assessment in the Appendix.

Although the 2002 Albania LSMS data were collected using a two-stage cluster sampling design, cluster identifiers are not available in the dataset used for this analysis. Consequently, we model municipalities as small areas and include area-level random effects only. With this specification, the covariance structure $\mathbf V_i$ captures dependence among units within areas through the area effect, but it does not explicitly model additional intra-cluster correlation arising from primary sampling units. If detailed cluster information were available, an extension of the NERHDP framework to incorporate cluster-within-area random effects would be a natural and practically relevant generalization, potentially improving variance estimation and predictive performance.

\begin{figure}[htbp]
        \centering
\includegraphics[width=0.6\textwidth, height=0.25\textheight]{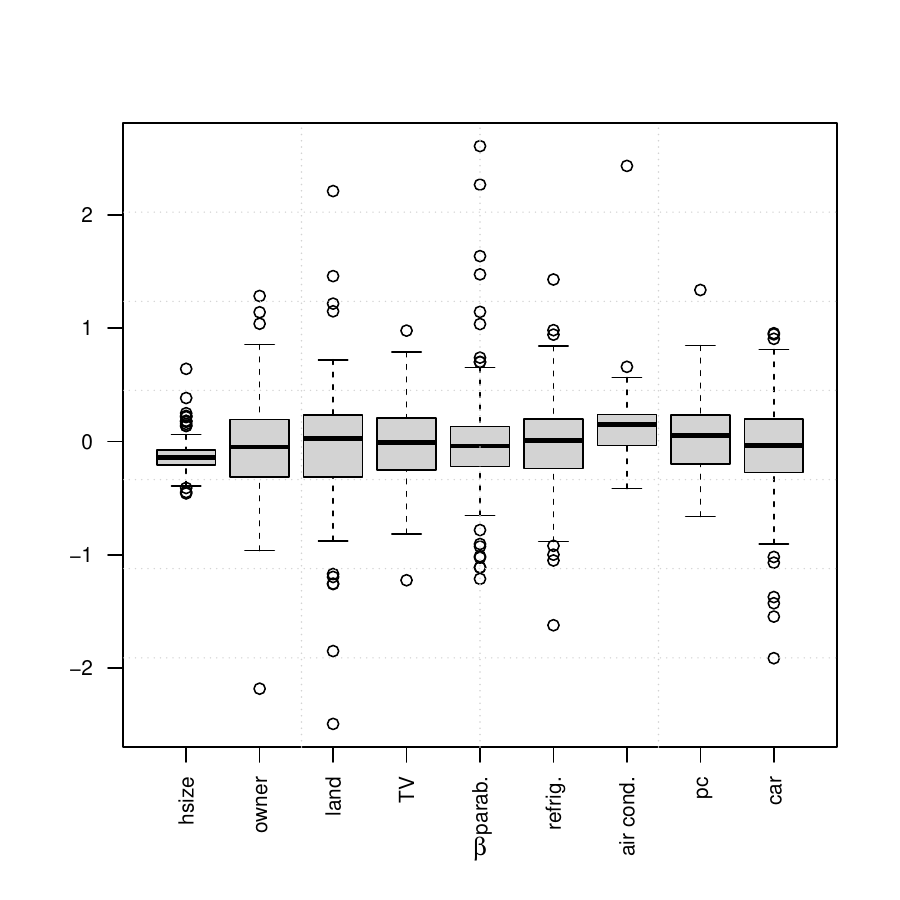}
        \caption{\centering Distributions of estimated regression coefficients fitted for each of the 213 municipalities of Albania.}
        \label{fig:beta}
\end{figure}

\begin{figure}[htbp]
        \centering
    \includegraphics[width=1\textwidth, height=0.25\textheight]{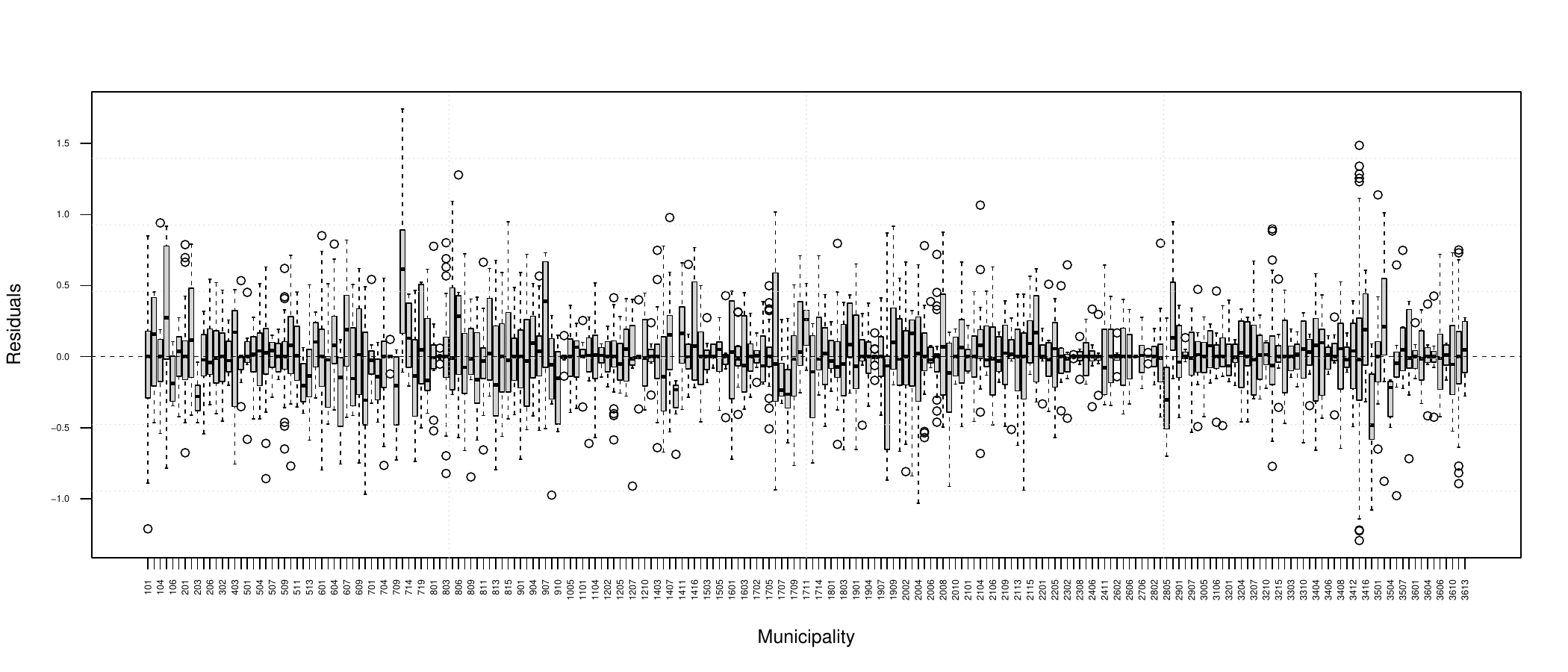}
        \caption{\centering Distributions of the standardized residuals by municipality.}
        \label{fig:res}
\end{figure}

To examine the existence of heterogeneity across municipalities, we fitted separate regression models between $Y_{ij}$ and the covariates for each municipality. Figures \ref{fig:beta} and \ref{fig:res} show boxplots of the estimated regression coefficients and standardized residuals, respectively. These results reveal potential variation across municipalities, providing empirical support for the use of a model that accommodates  heterogeneity. Additional diagnostic analyses, including assessments of normality, identification of outliers and high-leverage points, and tests for heterogeneity in regression coefficients, further support the suitability of the NERHDP model for this application. Details of these diagnostics are provided in the Appendix. Consistent with these findings, the model-based simulation results in Section \ref{sec:dbsim} further demonstrate that in the simulation setup, the CLS method under the NERHDP model performs well, showing (i) substantially lower bias and RRMSPE than the MR method when heterogeneity in regression coefficients and/or sampling variances exists, and (ii) improved computational efficiency relative to the algorithm proposed by \cite{lahiri2023nested}, making it more feasible for large-scale applications.

We apply the CLS method to obtain EBPs of HCRs and PGs by municipality, denoted as $\hat{F}_{0i}^{CLS}$ and $\hat{F}_{1i}^{CLS}$, respectively. The MSPEs of these FGT measures are estimated using the parametric bootstrap estimator $\text{mspe}_*(\hat{F}_{\alpha i}^{CLS})$ with $B = 100$ replicates. We then estimate the associated coefficients of variation (CVs) using:
\begin{eqnarray*}
\text{CV}(\hat{F}_{\alpha i}^{CLS}) = \{\text{mspe}_*(\hat{F}_{\alpha i}^{CLS})\}^{1/2}/\hat{F}_{\alpha i}^{CLS}. 
\end{eqnarray*}
Following \cite{molina2010small}, we also compute direct estimators of the FGT measures $\hat{F}_{\alpha i}^{w}$  and their variance estimates $\widehat{\text{Var}}_D(\hat{F}_{\alpha i}^{w})$ for sampled municipalities using weighted sample means (see Section \ref{sec:data}).

To assess the quality of our CLS estimator, we use a set of diagnostic tools based on the requirement that model-based small area estimates should align with the corresponding survey-weighted direct estimates, albeit more precise \citep{marino2019semiparametric}. Correlation coefficients between CLS and direct estimates for the HCR and PG are 0.796 and 0.802, respectively, suggesting strong agreement between the two approaches. Based on heuristic arguments, we adopt the goodness-of-fit diagnostic proposed by \cite{brown2001evaluation} to further evaluate coherence:
\begin{eqnarray*}
    W = \sum_i^{m_s}\frac{(\hat{F}_{\alpha i}^{w} - \hat{F}_{\alpha i}^{CLS})^2}{\widehat{\text{Var}}(\hat{F}_{\alpha i}^{w}) + \text{mspe}_*(\hat{F}_{\alpha i}^{CLS})}.
\end{eqnarray*}
With $W = 119.59$ for HCR and $W = 175.75$ for PG, both are below the 95th percentile of a $\chi^2$ distribution with 212 degrees of freedom ($\chi^2_{212, 0.95} = 177.39$), indicating that CLS estimates are coherent with direct estimates.

Table \ref{tab:albania} presents selected results across municipalities with varying sample sizes. The CLS estimators consistently exhibit lower estimated CVs than those of the direct estimators, even in areas with small samples. In several municipalities where direct estimates are zero, estimated CVs are undefined, highlighting a limitation of the direct estimation approach.

\begin{table}[htbp]
\centering
\caption{Population size, sample size, direct and CLS estimates of HCRs (\%) and PGs (\%), and the estimated CVs (\%) of direct and CLS estimators for the Albania municipalities
with sample size closest to minimum, first quartile, median, third quartile, and maximum.}
\label{tab:albania}
\begin{adjustbox}{width=1.0\textwidth}
\begin{tabular}{lccccccccccc}
  \toprule
 & & & \multicolumn{4}{c}{HCR}  &\multicolumn{4}{c}{PG}\\
 \cmidrule(lr){4-7} \cmidrule(lr){8-11}
Municipality & $N_i$ & $n_i$ & $\hat{F}_{0i}^w$ & $\hat{F}_{0i}^{CLS}$ & CV($\hat{F}_{0i}^w$) & CV($\hat{F}_{0i}^{CLS}$)& $\hat{F}_{1i}^w$ & $\hat{F}_{1i}^{CLS}$ & CV($\hat{F}_{1i}^w$) & CV($\hat{F}_{1i}^{CLS}$) \\ 
  \midrule
Hajmel & 1111 &    6 & 33.33 & 23.82 & 56.96 & 27.01 & 1.52 & 5.92 & 70.56 & 31.31 \\ 
Orenje & 1419 &   16 & 12.50 & 20.69 & 66.19 & 15.06 & 2.00 & 4.97 & 66.03 & 17.44 \\ 
B.Curri & 1488 &   40 & 20.00 & 24.54 & 32.82 & 8.57 & 5.65 & 5.51 & 43.01 & 9.56 \\ 
Gramsh & 2503 &   64 & 23.55 & 17.97 & 24.48 & 0.69 & 4.75 & 3.64 & 38.93 & 21.90 \\ 
Tirane & 89764 &  600 & 12.43 & 11.72 & 10.91 & 4.46 & 2.49 & 2.31 & 14.71 & 4.72 \\ 
   \hline
\end{tabular}
\end{adjustbox}
\end{table}

\begin{figure}[htbp]
    \centering
    \includegraphics[width=0.8\linewidth, height=0.25\textheight]{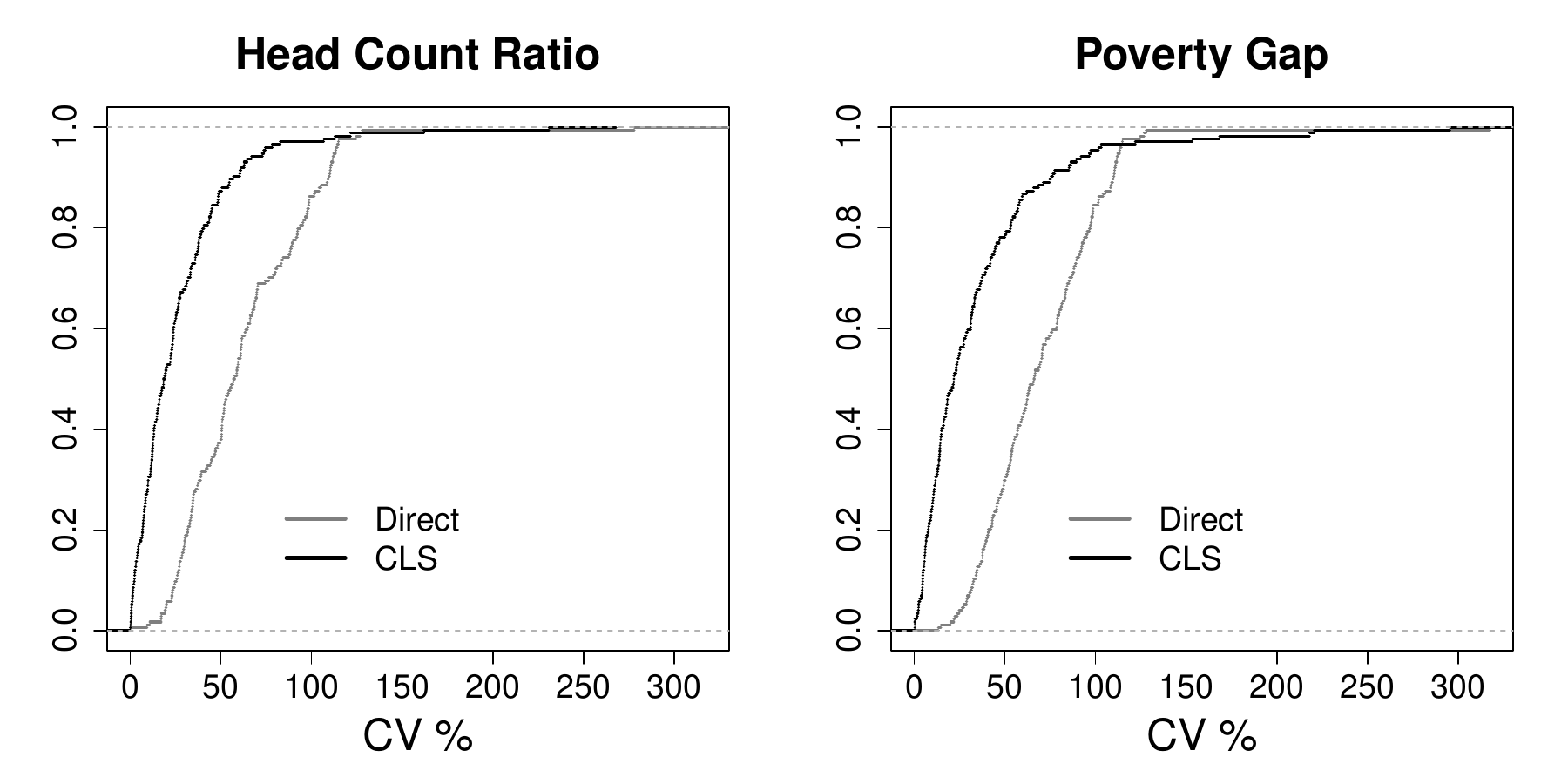}
    \caption{\centering CVs empirical cumulative density functions for the CLS and direct estimator.}
    \label{fig:ECDF}
\end{figure}

Figure \ref{fig:ECDF} compares the empirical cumulative distribution functions (ECDFs) of the estimated CVs for CLS and direct estimators. The ECDFs of the CLS estimators dominate those of the direct estimators for both HCR and PG, illustrating their improved precision. Notably, about 78\% (HCR) and 89\% (PG) of municipalities have direct estimate with estimated CVs exceeding the conventional 33\% reliability threshold in small area estimation \citep{marino2019semiparametric}. In contrast, these percentages drop to 28\% and 34\% under the CLS approach.

\begin{figure}[htbp]
    \centering
    \includegraphics[width=0.6\textwidth, height=0.25\textheight]{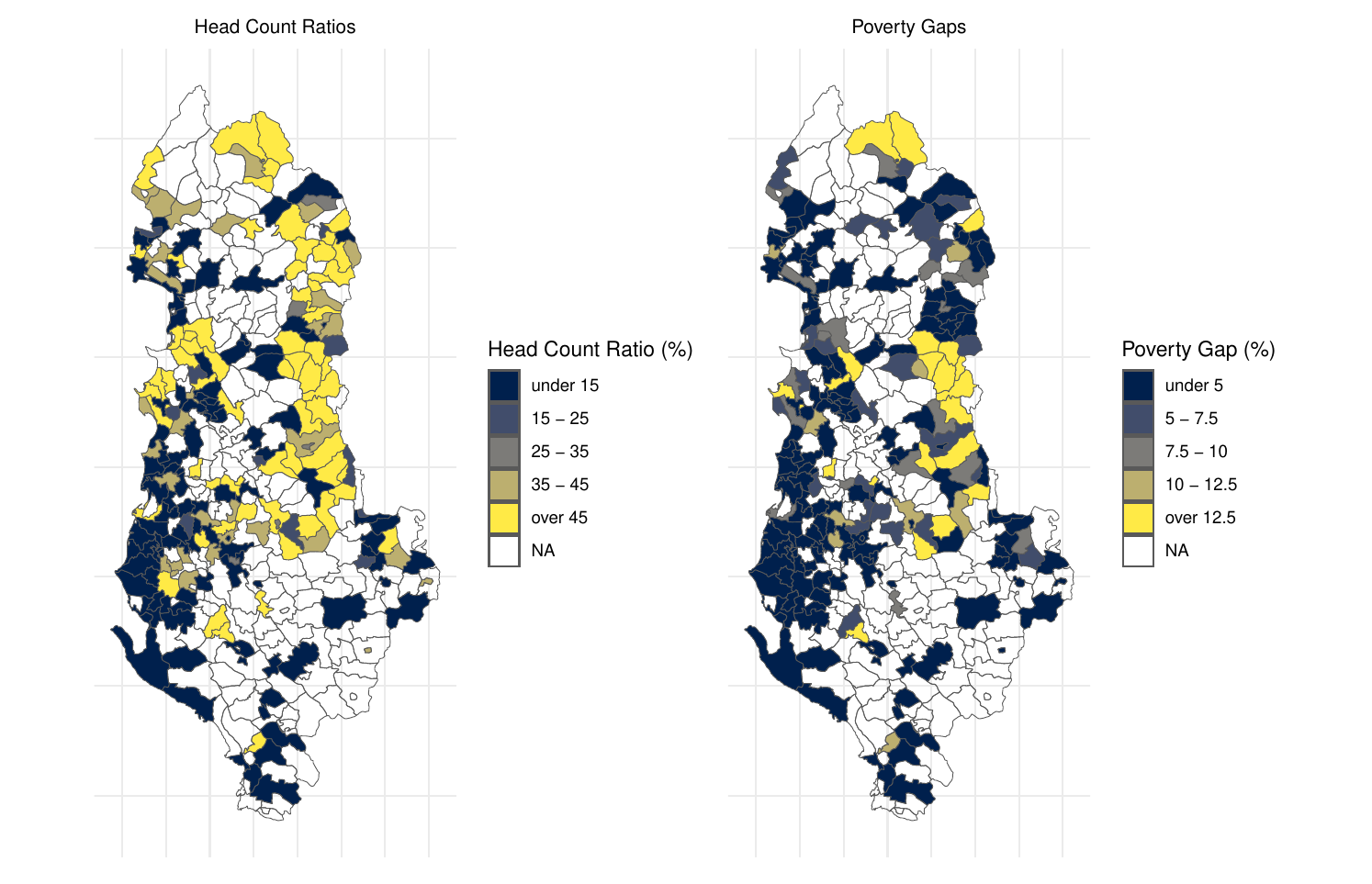}
    \caption{\centering Municipality-level direct estimates of HCRs and PGs in Albania.}
    \label{app:dirmap}
\end{figure}

\begin{figure}[htbp]
    \centering
     \includegraphics[width=0.6\textwidth, height=0.25\textheight]{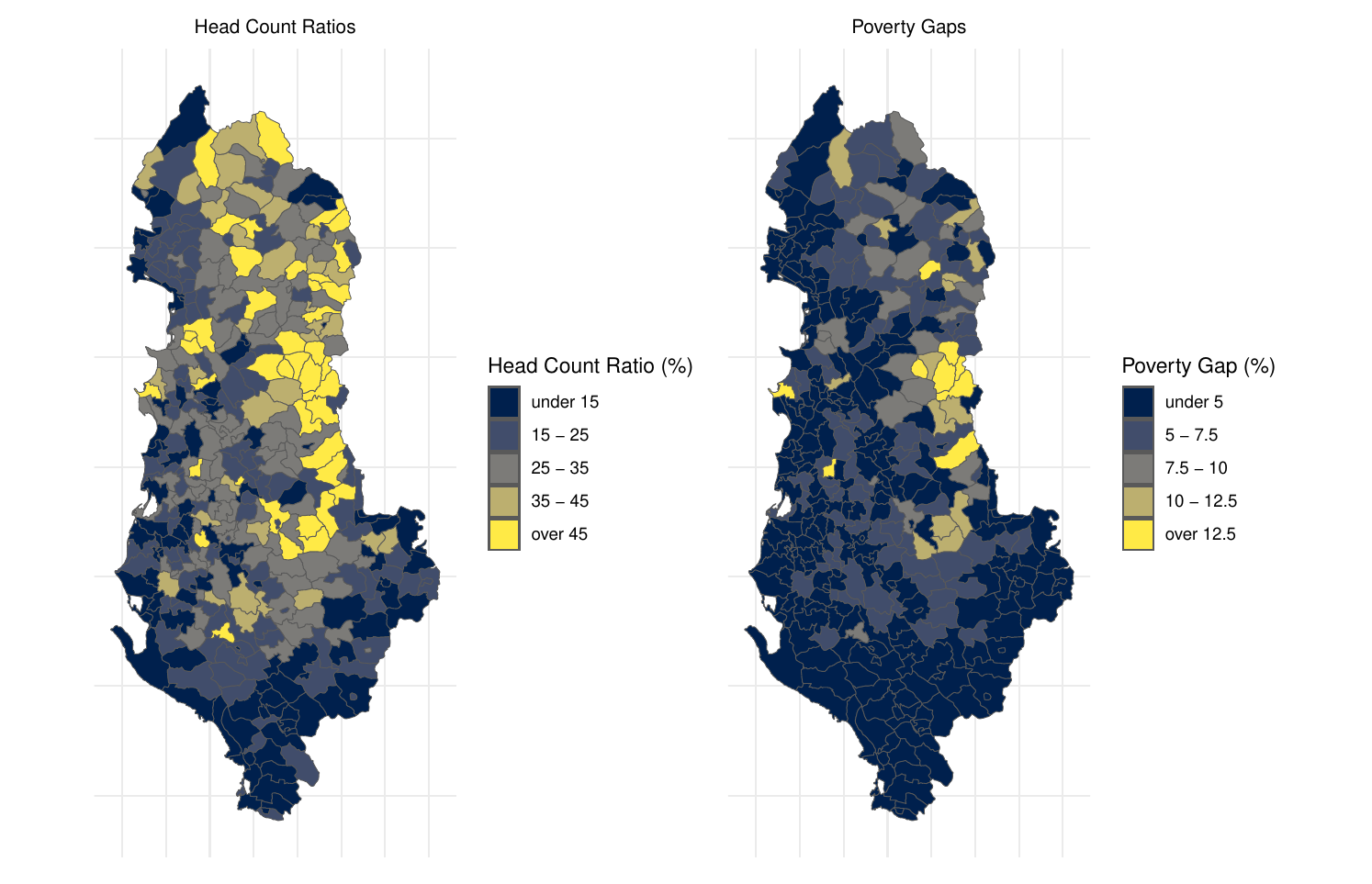}
    \caption{\centering Municipality-level CLS estimates of HCRs and PGs in Albania.}
    \label{app:Albania}
\end{figure}

Figures \ref{app:dirmap} and \ref{app:Albania} present municipality-level poverty maps for Albania using direct and CLS estimates, respectively. The white areas in Figure \ref{app:dirmap} represent regions where direct poverty estimates are unavailable due to zero sample size, underscoring a key limitation of the direct estimation approach in small-area analysis. In contrast, Figure \ref{app:Albania} presents poverty maps generated using our proposed CLS method, which provides more comprehensive and reliable estimates across all municipalities.  The CLS maps indicate that municipalities in the Bulqize district, such as Ostren, Gjorice, Shupenze, and Bulqize, exhibit the highest HCR and PG values, significantly exceeding those observed in other municipalities. Similarly, municipalities like Gjoaaj in the Peqin district, Katund I in the Durres district, and Arren in the Kukes district also show relatively high poverty levels, highlighting the need for targeted policy interventions to improve living conditions in these areas. In contrast, municipalities in the southern regions display substantially lower poverty levels, reflecting better economic conditions and higher living standards. 
Although no formal field validation study is available for these municipality-level SAE results, the spatial patterns obtained here are broadly consistent with earlier poverty mapping analyses for Albania \citep{betti2003poverty,tzavidis2008m}, providing indirect empirical support for their plausibility.

Overall, the CLS method provides a more complete and reliable representation of poverty distribution in Albania. By producing estimates for all municipalities, it addresses the data limitations of direct estimation and enables a more comprehensive understanding of regional poverty disparities. The identification of high-poverty areas in the northern and central regions highlights the need for geographically targeted poverty alleviation programs, while the lower poverty levels in the south suggest regional inequalities that should inform national development strategies. The robustness of the CLS estimates, even in municipalities with small sample sizes, suggests that, when the data structure and underlying conditions are comparable to those considered in our simulation studies, our method may offer practical value for policy-oriented poverty mapping.

\section{Discussion}\label{sec:discussion}
The proposed model bridges the gap between existing approaches that either assume identical model parameters across areas, potentially leading to misspecification, or rely on random slope or sampling variance specifications that demand strong distributional assumptions. By leveraging area-specific estimating equations and appropriately constructed residuals, we offer a framework for deriving empirical best prediction of complex poverty measures that captures variability across areas without the instability associated with fixed effects in small samples. On the computational side, we introduce a novel, computationally efficient, and data-driven procedure for estimating model parameters, which is critical for practical applications. Additionally, we develop a new method to estimate tuning parameters for out-of-sample areas, enabling the derivation of both area-specific regression coefficients and sampling variances, even for out of sample areas. 

Based on the Albania data, our simulation studies highlight the advantages of the proposed CLS method in prediction accuracy, which outperforms the MR method in terms of both bias and RRMSPE, when the assumptions of constant regression coefficients and/or sampling variances across areas are violated. These results demonstrate the effectiveness of our approach in handling heterogeneous and complex data structures. It is important to note, however, that no single SAE method is universally optimal. The relative performance of NERHDP, traditional NER-based EBP, or alternative approaches such as ELL depends on the underlying data structure, the degree of heterogeneity across areas, model specification, and the availability and quality of auxiliary information. The proposed method is particularly advantageous when regression coefficients and/or sampling variances vary across areas, but under homogeneous settings simpler models may perform equally well or better. We therefore encourage future research to further assess the methodology under a broader range of simulation scenarios.

It is important to note that estimates for out-of-sample areas, whether derived from CLS or MR, remain synthetic due to the absence of direct observations. Although the proposed method performs well in out-of-sample areas when strong auxiliary information is present, future research on reducing the synthetic nature of these estimates is warranted.

In this paper, we empirically evaluated the performance of the proposed model parameter estimators and small area predictors.  Investigation of asymptotic properties will be an interesting future research topic, but for such research we  would need a suitable theoretical framework that is likely to  require large area specific sample sizes.  We did not explore such asymptotic  theory in this paper because in our applications the assumption of large area specific sample sizes is not realistic. Future work could also investigate formal inference procedures, more detailed structures for the tuning parameter $\tau_i$ and model selection strategies for the NERHDP model, since these components are essential for practical implementation.

While the proposed NERHDP model demonstrates promising results based on the Albania data in reducing bias and improving precision, its performance may vary under different data structures and sampling conditions. For instance, under complex survey design, such as multi-stage sampling conducted, with available cluster identifiers and rich contextual variables, alternative modeling strategies may better capture the data structure and yield improved performance.

In our simulation studies, EBP-type approaches tend to outperform the ELL-type predictor. However, in practice, the relative performance of EBP- and ELL-type methods depends critically on the underlying data structure, the availability of auxiliary information at different levels of aggregation, and the specification of the variance structure. As highlighted in recent review articles (e.g., \citealp{molina2022estimation}; \citealp{das2024small}), neither class of methods is uniformly superior. ELL-type estimators may benefit from rich contextual information and flexible variance modeling, whereas EBP-type predictors may perform well when the nested error regression framework adequately represents the data-generating mechanism.

Open questions also remain regarding the estimation of variance components, particularly for random effects. Although no zero or negative variance estimates were observed in our study, the algorithm does not guarantee positivity of variance estimates of random effects. Non-positive variance estimates would compromise the bootstrap-based uncertainty measures proposed here. Further research is therefore needed to ensure valid variance estimation and enhance the robustness of the NERHDP framework. 

Finally, similar to much of the existing literature on empirical best prediction in the poverty mapping literature, this paper assumes normality for both the random effects and the error terms. Nevertheless, we acknowledge that normality is not a prerequisite for the development of EBP methodology. Our empirical findings suggest that future research should explore alternative parametric distributions motivated by observed data characteristics, thereby improving model flexibility and robustness in practical applications.

\appendix

\section*{Appendix: Diagnostic checks for Albanian application}
While a comprehensive treatment of model diagnostics is beyond the scope of the current paper, we acknowledge the importance of assessing model adequacy in small area estimation contexts. We conducted several diagnostics and discussion to support the plausibility of the usage of NERHDP model in the Albanian application. 
\begin{description}
    \item[1. Normality of residuals: ] To assess the plausibility of the normality assumption underlying the nested error regression framework, we fitted a two-level linear mixed model to the Albanian data and examined the empirical distributions of both area-level random effects and unit-level residuals. Normal probability plots are displayed in Figure 9. The Shapiro–Wilk tests yield p-values of $2.29 \times 10^{-10}$ at level 1 and $2.78 \times 10^{-6}$ at level 2, indicating departures from strict normality.\\ These results suggest that normality should be viewed as an approximation rather than a literal description of the data-generating process. Such deviations are not unexpected in welfare and expenditure data, even after logarithmic transformation. Within the superpopulation framework adopted here, normality is primarily a working modeling assumption that facilitates derivation of conditional distributions required for empirical best prediction. \\Importantly, the proposed NERHDP approach incorporates robustness features through influence-function-based estimating equations (with Huber-type weights), which mitigate the impact of outliers and moderate departures from distributional assumptions. Moreover, empirical best predictors under nested error regression models are known to be relatively stable under mild deviations from normality, particularly when the primary goal is conditional prediction rather than exact likelihood-based inference.\\Taken together, these diagnostics motivate the use of a flexible and robust modeling strategy rather than strict reliance on Gaussian assumptions. Future work may consider semiparametric or distributionally robust extensions of the conditional prediction framework to further relax Gaussian assumptions.
    \begin{figure}[htbp]
        \centering
        \includegraphics[width=0.8\linewidth]{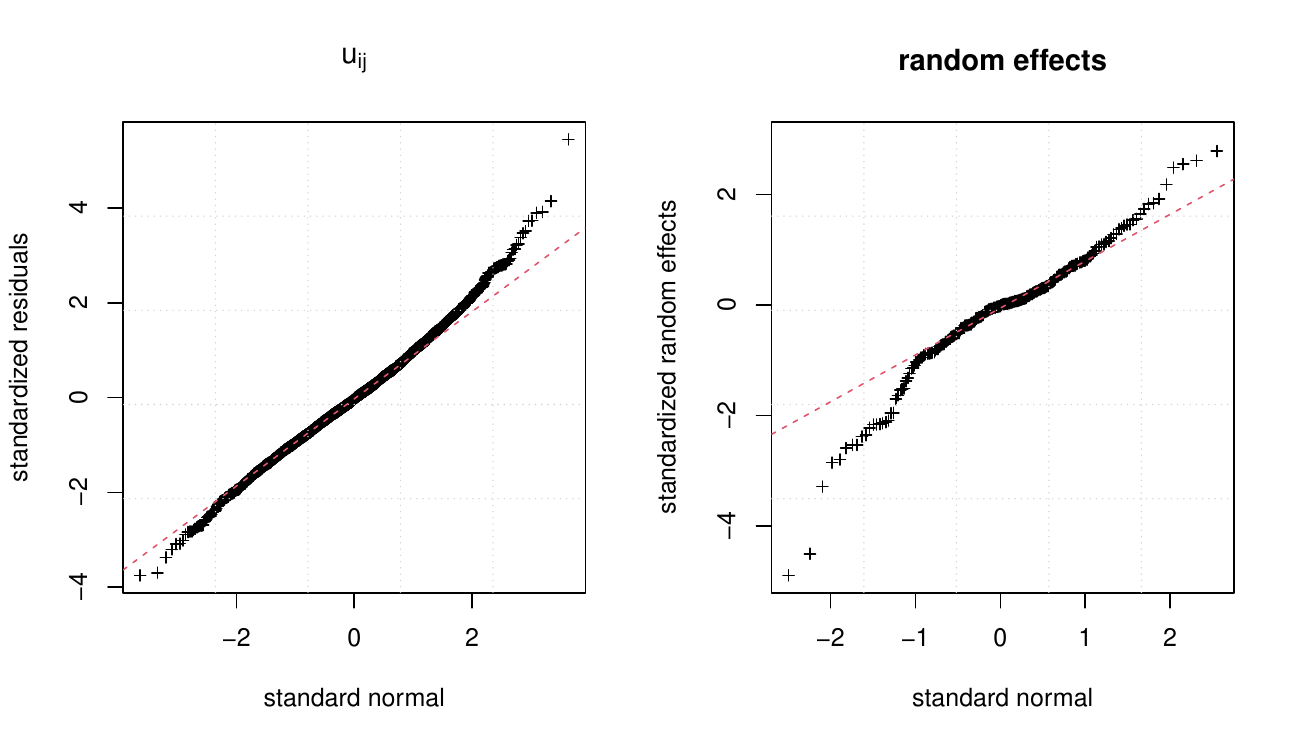}
        \caption{\centering Normal probability plots of level 1 (left) and level 2 residuals (right) derived by fitting a two level linear mixed model to sample data.}
        \label{fig:normality}
    \end{figure}
    \item[2. Outliers and high leverage points: ] Following \cite{zewotir2007unified}, we examined potential outliers and high leverage points in the sample data. Specifically, the diagonal elements $s_{ii}$ of the matrix $\mathbf{S} = \sigma^2_{\epsilon}\mathbf{P}$, with $\mathbf{P} = \mathbf{V}^{-1} - \mathbf{V}^{-1}\mathbf{X}(\mathbf{X}^\prime \mathbf{V}^{-1}\mathbf{X})^{-1}\mathbf{X}^\prime \mathbf{V}^{-1}$, are used to identify high leverage points, where $\mathbf{X}$ is the design matrix and $\mathbf{V}$ is the covariance matrix of the linear mixed model. The diagnostic involves plotting $s_{ii}$ against $\hat{e}^2_{i}/\hat{\mathbf{e}}^\prime\hat{\mathbf{e}}$, where $\hat{e}_{i}$ denotes the EBLUP residual. Points are expected to cluster around the upper-left corner of the plot. Those in the lower-left corner (small $s_{ii}$) indicate high leverage, while those on the right (large relative squared residuals) suggest outliers. As shown in Figure~\ref{fig:outlier}, both outliers and high leverage points are potentially present.
    \begin{figure}[htbp]
        \centering
        \includegraphics[width=0.8\linewidth]{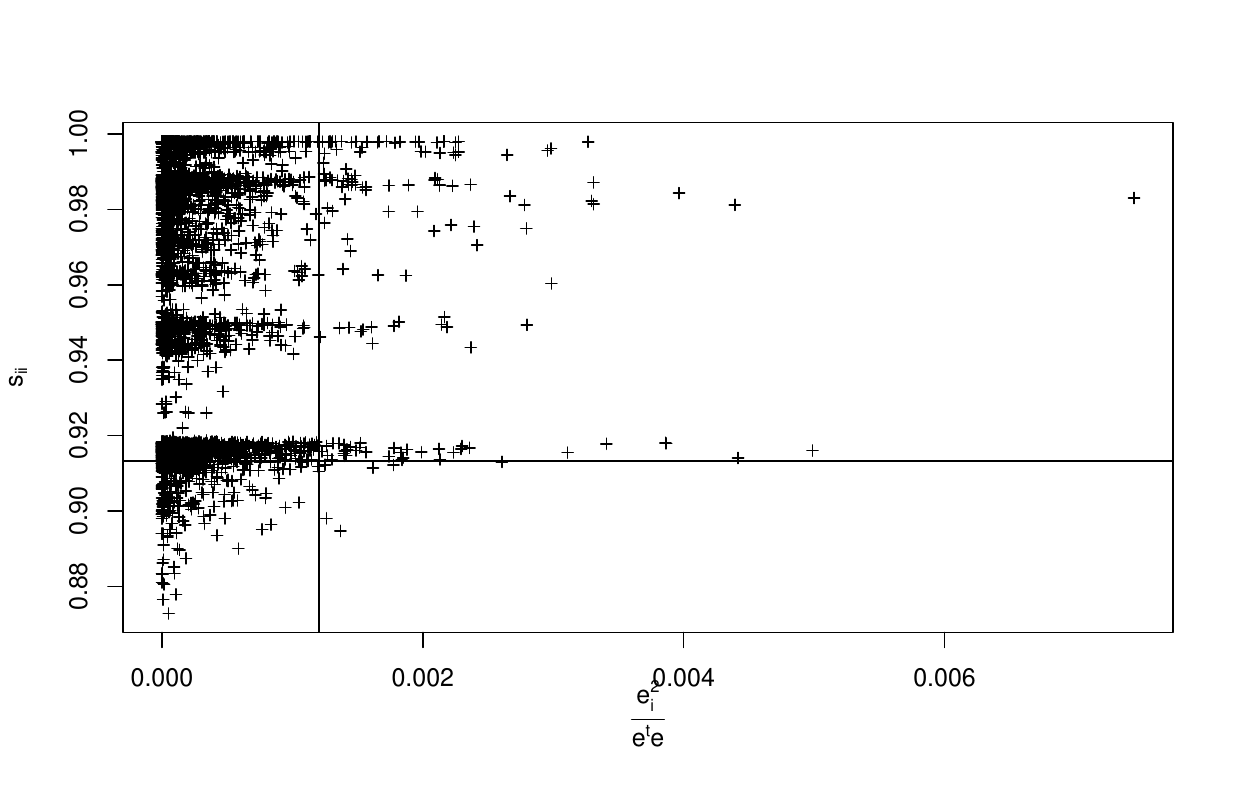}
        \caption{\centering Plot to detect both outliers and high leverage points: $s_{ii}$ versus $\hat{e}^2_{i}/\hat{\mathbf{e}}^\prime\hat{\mathbf{e}}$}
        \label{fig:outlier}
    \end{figure}
These diagnostics highlight the need for a more robust and flexible approach. Accordingly, we fitted the NERHDP model to the Albanian data, setting the tuning constant in the Huber influence function to $c = 1.345$.
    \item[3. ] In addition, we conducted Hausman test \citep{hausman1978specification} to assess the heterogeneity of regression coefficients. The test statistic is defined as: $h_i = (\hat{\bm\beta} - \hat{\bm\beta}_i)^\prime\bigg(\widehat{Var}(\hat{\bm\beta}) -\widehat{Var}(\hat{\bm\beta}_i)\bigg)^{-1}(\hat{\bm\beta} - \hat{\bm\beta}_i)$ with $\widehat{Var}(\hat{\bm\beta}_i) = (\sum_{l=1}^{m}\mathbf{X}_l\hat{\mathbf{V}}^{-1}_{l;i}\mathbf{X}_l^\prime)^{-1}$.  Under the null hypothesis that regression coefficients are homogeneous across small areas, this statistic approximately follows a $\chi^2_{(p-1)\times m}$ distribution. The observed value of 6991.664 yields a very small $p$-value, leading us to reject the null hypothesis. This provides strong evidence of heterogeneity in the regression coefficients and aligns with the boxplots in Figure 8, further supporting the use of the NERHDP model for the Albanian data.
\end{description}

\bibliographystyle{abbrvnat}
\bibliography{ref}

\end{document}